\documentstyle[galley,graphicx,epsf]{mn} 
\begin{document}
\title{Five old open clusters more in the outer Galactic disk}
\author[G. Carraro et al.]        
{Giovanni Carraro$^{1,2}$, Yuri Beletsky$^3$, Gianni Marconi$^1$
\thanks{email: gcarraro,gmarconi@eso.org, beletsky@lco.cl}\\ 
$^1$European Southern Observatory, Alonso de Cordova 3107
Vitacura, Santiago de Chile, Chile\\
$^2$Dipartmento di Astronomia, Universit\'a di Padova, Vicolo Osservatorio 3, I-35122, Padova,
Italy\\
$^3$Las Campanas Observatory, Carnegie Institution of Washington, Colina el Pino, Casilla 601,
La Serena, Chile \\
} 
 
\date{\it Submitted: June 2011} 
\pubyear{2010} 
%
%
\maketitle 
\label{firstpage}
            
\begin{abstract} 
New photometric material is presented for 6 outer disk supposedly old,  Galactic star clusters:
Berkeley~76, Haffner~4, Ruprecht~10, Haffner~7, Haffner~11, and
Haffner~15, that are projected against the rich and complex Canis Major overdensity at $225^o
\leq l \leq 248^o $, $-7^o \leq b \leq  -2^o$.  
This CCD data-set, in the UBVI pass-bands, is used to derive their fundamental parameters, in particular
age and distance. 
Four of the program clusters turn out to be older than 1  Gyr. This fact makes them ideal targets
for future spectroscopic campaigns aiming at deriving their metal abundances.
This, in turn, contributes to increase the number of well-studied outer disk old open clusters.  
Only Haffner~15, previously considered an old cluster, is found to be a young, significantly reddened
cluster, member of the Perseus arm in the third Galactic quadrant.
 As for Haffner~4, we suggest an age of about half a Gyr.
The most interesting result we found is  that Berkeley~76 is probably located at more than
17 kpc from the Galactic center, and therefore is among the most peripherical old open clusters 
so far detected. 
Besides, for Ruprecht~10 and Haffner~7, which were never studied before, we propose ages larger than
1 Gyr.\\
All the old clusters of this sample are scarcely populated and show evidence of tidal
interaction with the Milky Way, and are therefore most probably in advanced stages of 
dynamical dissolution.
\end{abstract}

 
\section{Introduction}
The structure and evolution of the outer Galactic disk has been the subject
of intense investigation in the last years
(Frinchaboy et al. 2004, Momany et al. 2006, Carraro et al. 2007a, 2010). This is because the outer disk 
in the anti-center direction is less dominated by confusion than other Galactic directions,
a fact which renders it easier to describe its structure, and to search for 
signatures of ongoing or past accretion events. 
The latter consist of the Monoceros Ring (Chou et al. 2010 and references therein),
and the Canis Major overdensity (Momany et al. 2004). However,  
both these structures are now considered by most researchers to be caused
by the warped and flared Galactic disk (L\'opez-Corredoira et al. 2007, 2012; Carraro et al. 2008).\\ 

\begin{figure*}
\includegraphics[width=0.45\hsize]{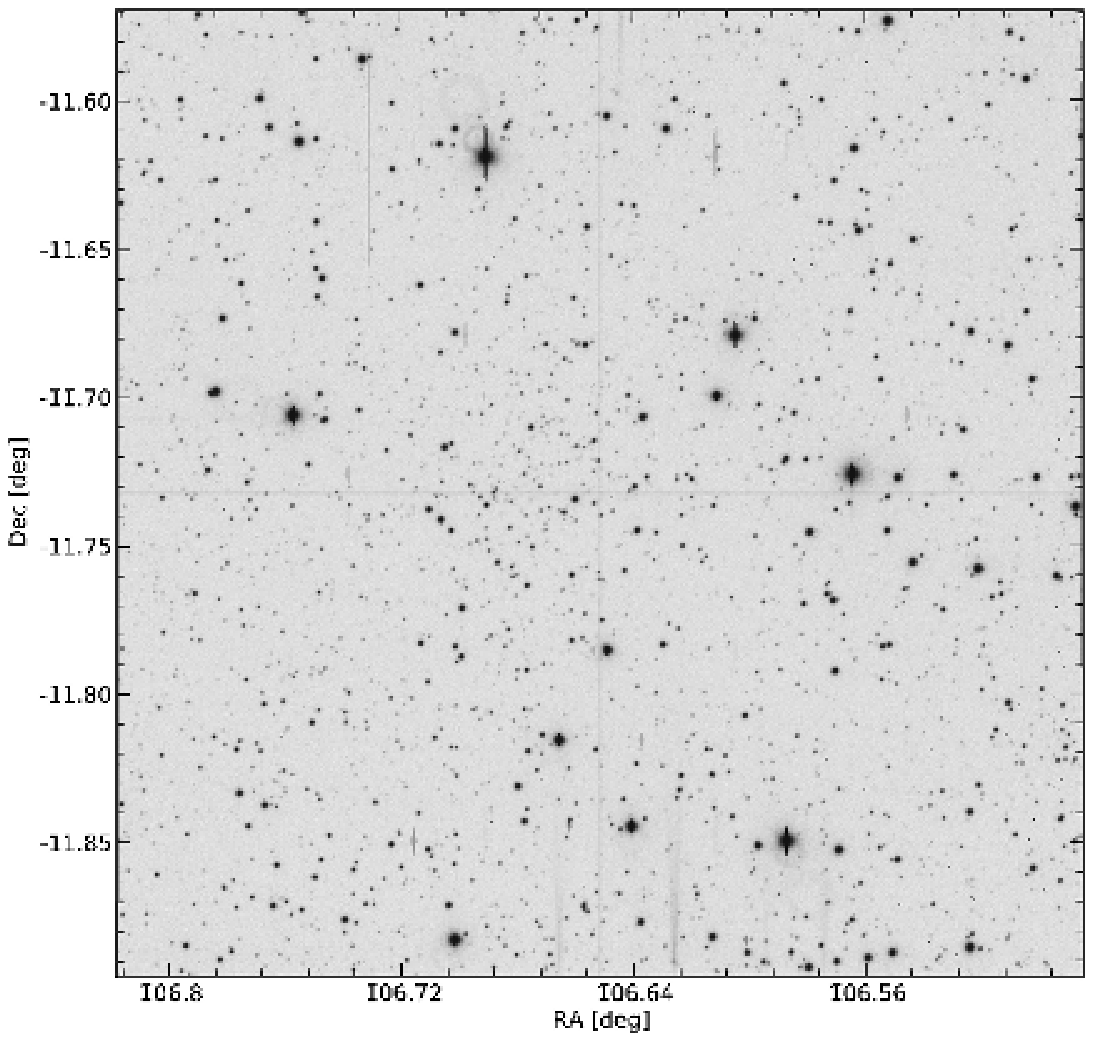}
\includegraphics[width=0.45\hsize]{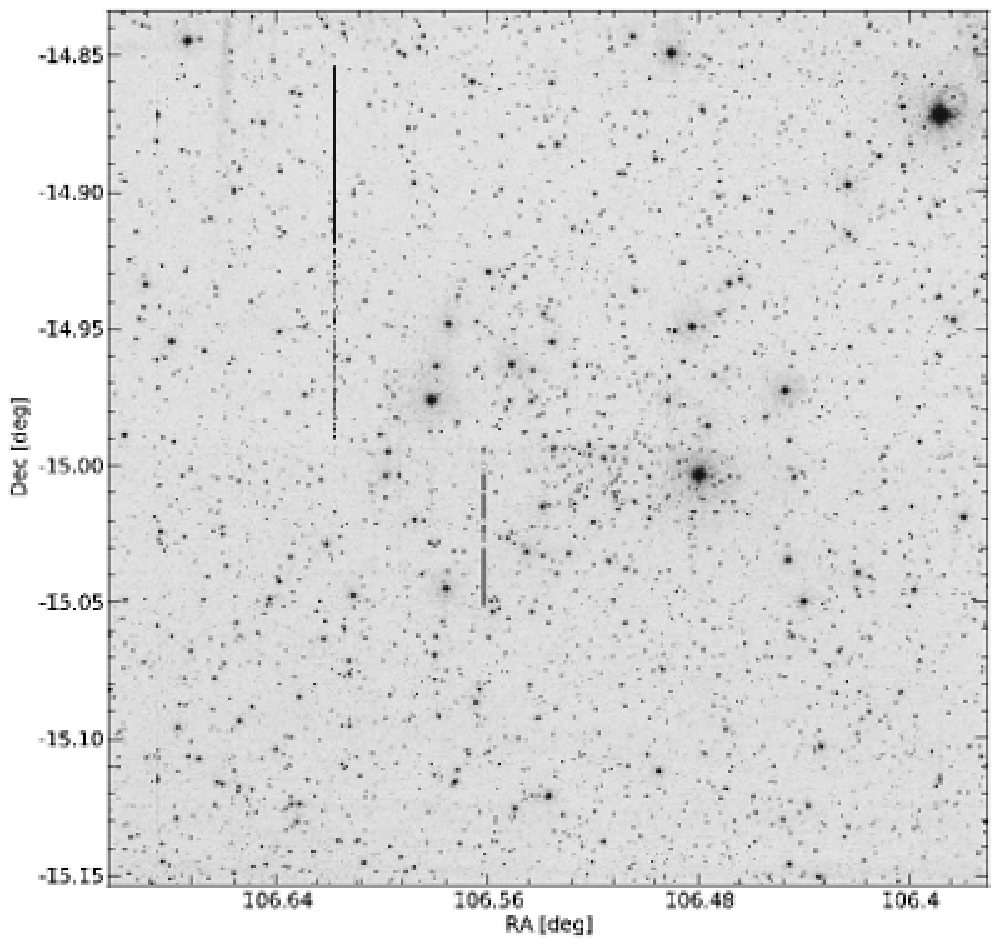}
\includegraphics[width=0.45\hsize]{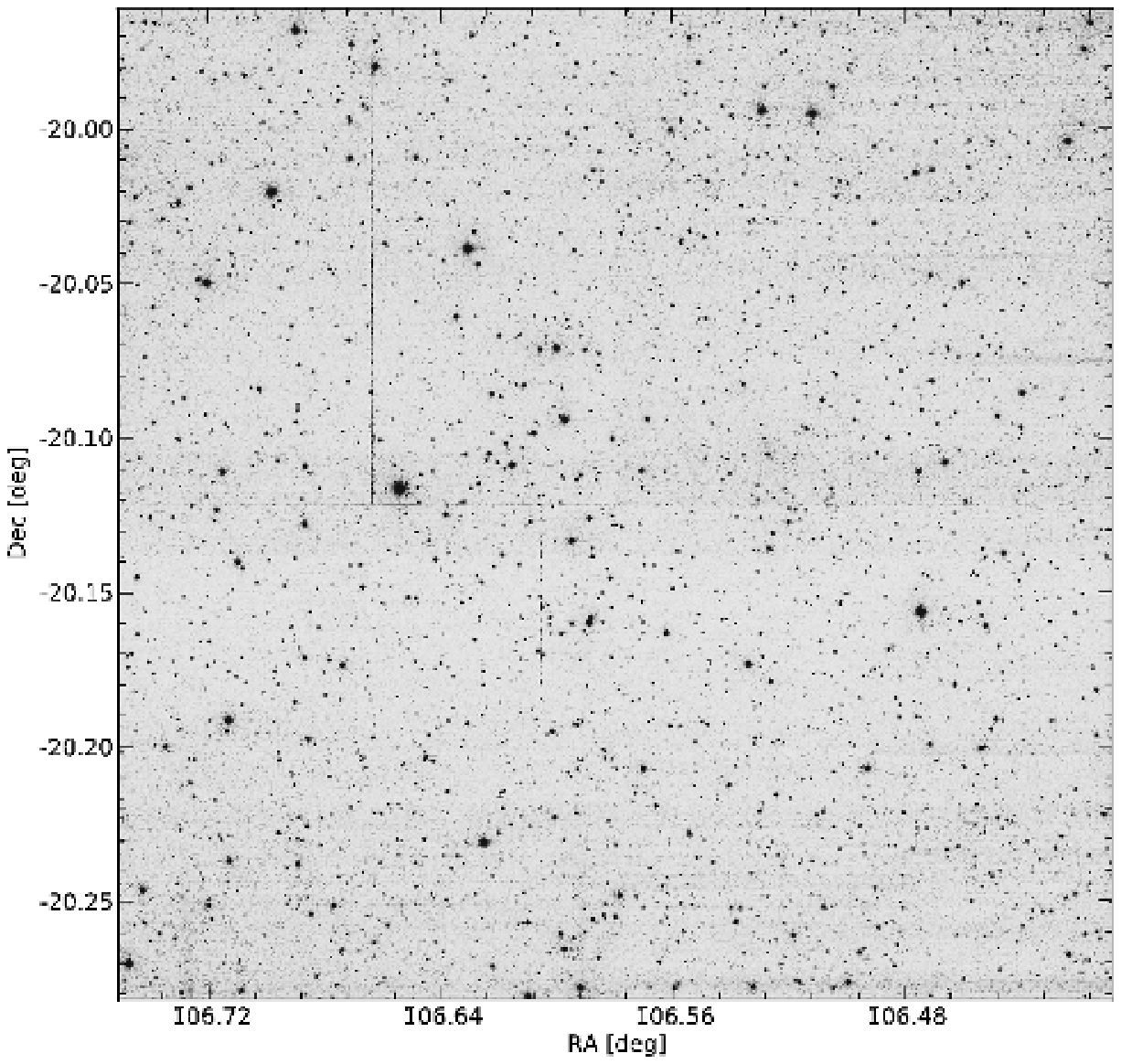}
\includegraphics[width=0.45\hsize]{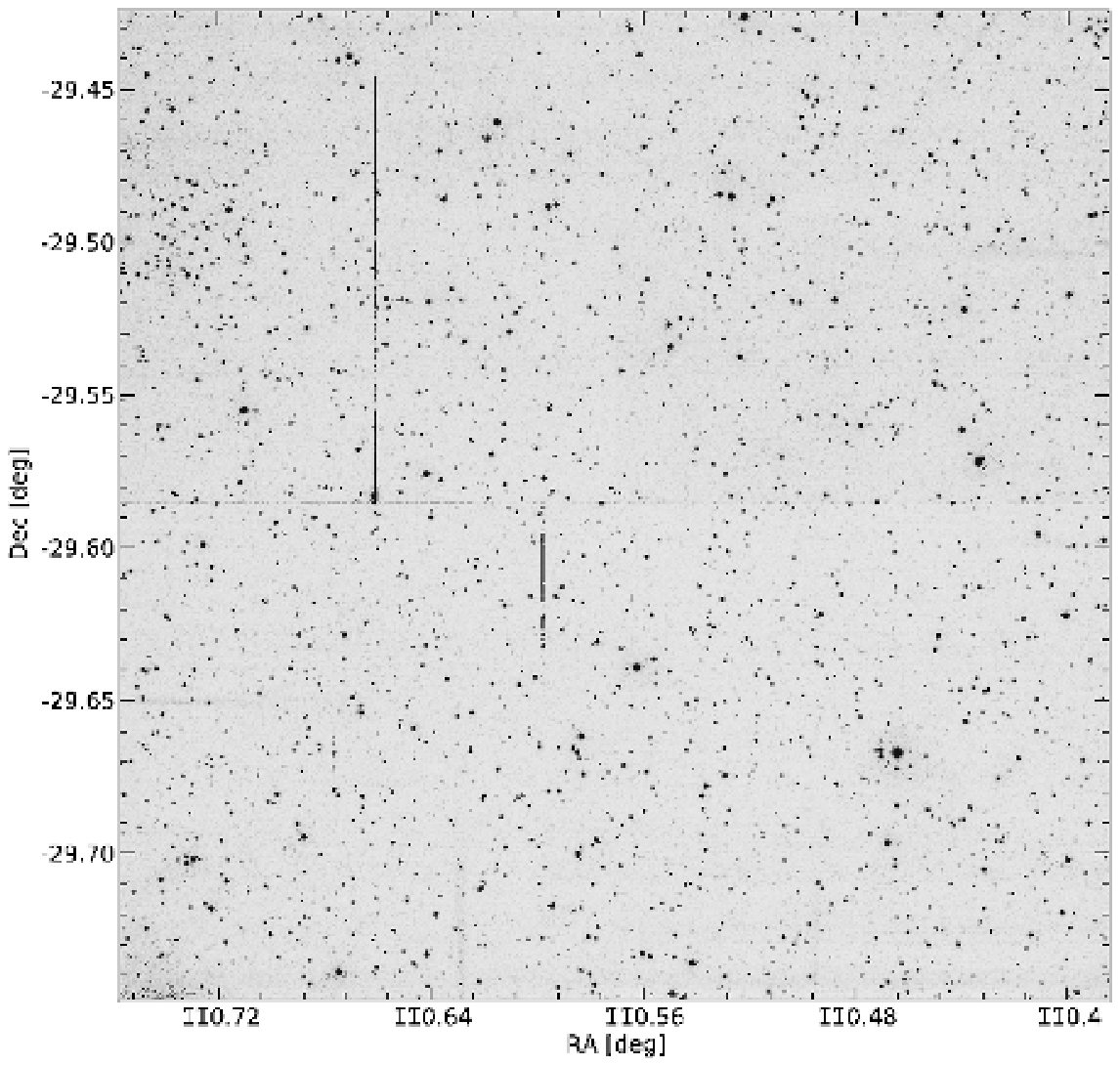}
\includegraphics[width=0.45\hsize]{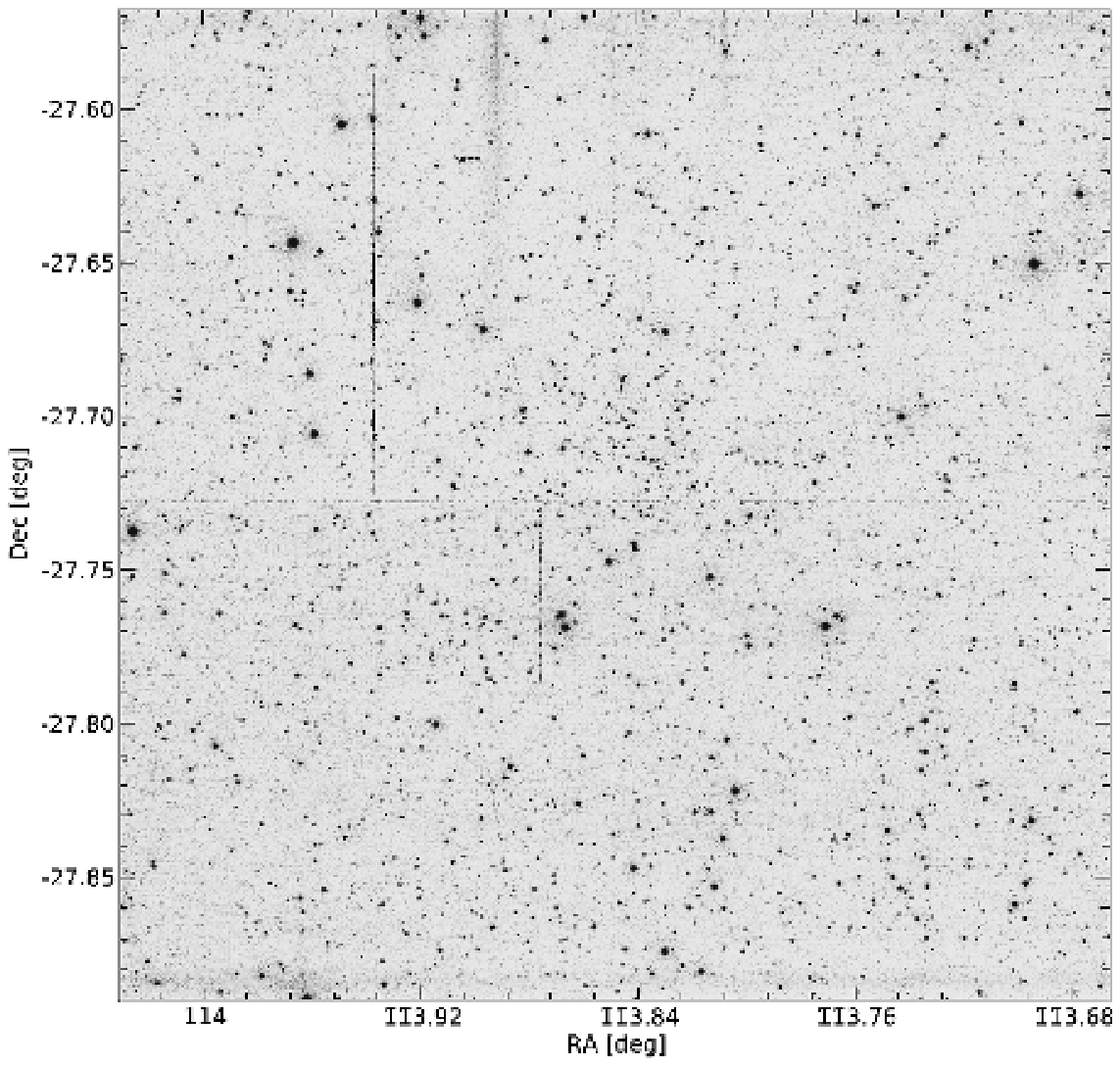}
\includegraphics[width=0.45\hsize]{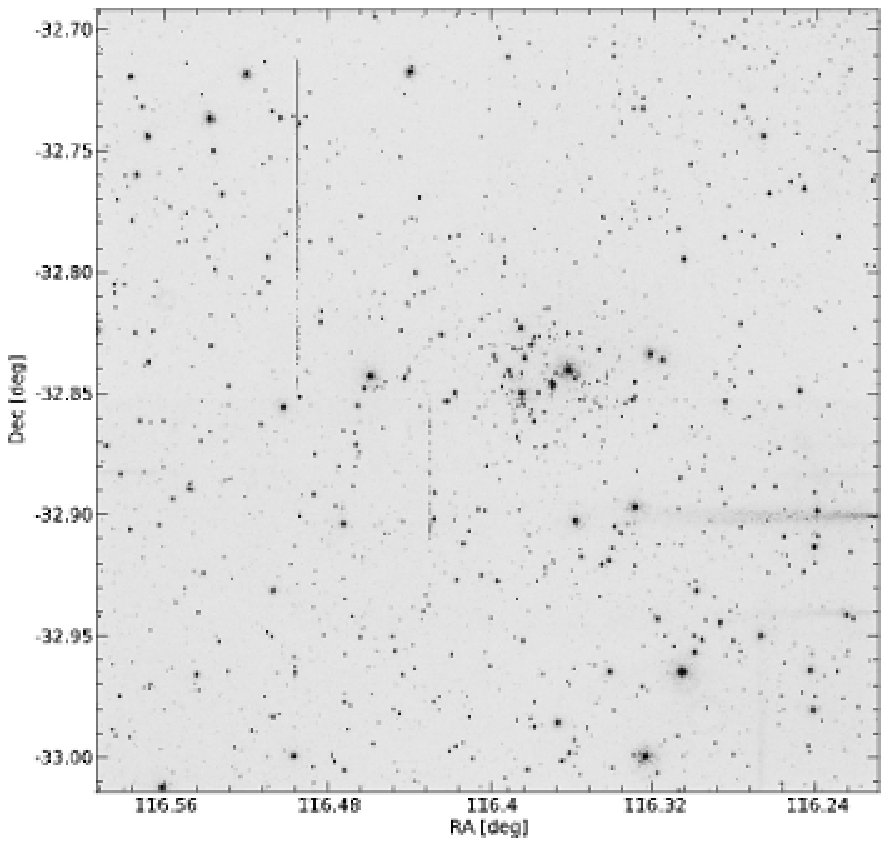}
\caption{Montages of all CTIO CCD frames for the clusters under study. 
{\it Top raw} (from left to right)-   Berkeley~76, Haffner~4;
{\it middle raw} (from left to right) -  Ruprecht~10,
Haffner~7; {\it bottom raw}(from left to right) -  Haffner~11, Haffner~15.}
\end{figure*}

\begin{table*}
\tabcolsep 0.2truecm
\caption{Basic parameters of the clusters under investigation.
Coordinates are for J2000.0. We list also equatorial coordinates in degrees (column 4 and 5), 
to help the reader to look at Figs.~1 and 3. The last column reports the expected reddening along the line
of sight all the way to infinity
according to Schlegel et al. (1998) maps.}
\begin{tabular}{cccccccc}
\hline
\hline
\multicolumn{1}{c}{Name} &
\multicolumn{1}{c}{$RA$}  &
\multicolumn{1}{c}{$DEC$}  &
\multicolumn{1}{c}{$RA$}  &
\multicolumn{1}{c}{$DEC$}  &
\multicolumn{1}{c}{$l$} &
\multicolumn{1}{c}{$b$} &
\multicolumn{1}{c}{$E(B-V)_{\infty}$}\\
\hline
& {\rm $hh:mm:ss$} & {\rm $^{o}$~:~$^{\prime}$~:~$^{\prime\prime}$} & [deg] & [deg] & [deg] & [deg] & mag\\
\hline
Berkeley 76  & 07:06:24 & -11:37:00 & 106.60 & -11.62 & 225.099  & -1.998 & 0.740\\
Haffner 4    & 07:06:12 & -14:59:00 & 106.55 & -14.98 & 227.940  & -3.586 & 0.721\\
Ruprecht 10  & 07:06:25 & -20:05:00 & 106.60 & -20.08 & 232.553  & -5.854 & 0.645\\
Haffner 7    & 07:22:55 & -29:30:00 & 110.73 & -29.50 & 242.673  & -6.804 & 0.270\\
Haffner 11   & 07:35:25 & -27:43:00 & 113.85 & -27.72 & 242.395  & -3.544 & 0.824\\
Haffner 15   & 07:45:32 & -32:51:00 & 116.38 & -32.85 & 247.952  & -4.158 & 1.558\\
\hline\hline
\end{tabular}
\end{table*}

\noindent
Among the various tracers routinely used to probe the outer disk, Galactic open
clusters offer the advantage that it is relatively easy to obtain estimates of their ages and distances, and that they are 
ubiquitous in the outer disk. This is a valid statement for both old and young open clusters, since the deficiency
of old open clusters is most visible in the inner disk, due to the more difficult enviroment which prevents a cluster
to survive fo a long time.\\

\noindent
In this paper we derive basic parameters of six anticenter clusters, located in the third quadrant of the Galactic disk:
Berkeley~76, Haffner~4, Ruprecht~10, Haffner~7, Haffner~11, and Haffner~15 
(see Table~1, where for each cluster we report
Equatorial and Galactic coordinates, together with an estimate of the reddening at infinity in their direction).
The rich fields they are projected against have been already studied in detail in V\'azquez et al. (2008), where
we highlight the properties of the conspicuous young populations we detected and use it to probe the spiral
structure in that quadrant.
The analysis of the clusters we performed in that paper was quite preliminary, and for this reason in this paper we are going
to present the data, show the photometric diagrams, provide an extensive analysis of them, 
and propose updated estimates of distance and age for these clusters. In a few cases, these estimates are 
given for the very first time.\\

\noindent
Interestingly enough, we anticipate that these clusters turn out to be all older than 1 Gyr, except for
Haffner~5, which is a young, highly obscured cluster. This
constitutes quite a useful result {\it per se}, since it contributes to improve the statistics of old clusters in the Galactic
disk outer regions and therefore can help to provide tighter constraints for chemical evolution models aiming at recovering
the chemical history and assembly of the outer disk (Carraro et al. 2007a).\\

\noindent
The paper is organized as follows:  In Section~1 we present the data, the observation strategy,
and the extraction and calibration of the photometry. In the same Section, details are provided on
the astrometry, completeness, and star count analysis. A summary of the information available in the literature
for these clusters is presented in Section~2, while Section~3 introduces the basic diagrams we use
to extract clusters' fundamental parameters. Finally, Section~4 summarizes the results and discusses their
implications.

  \begin{figure}
   \centering
   \includegraphics[width=\columnwidth]{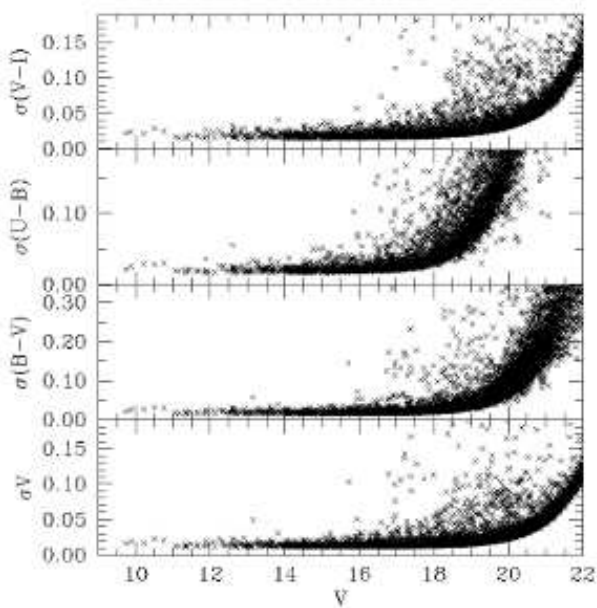}
   \caption{Trend of photometric in V, B-V, U-B and V-I as a function of V.}
    \end{figure}

\subsection{Observations and data reduction}       
The targets were observed with the Y4KCAM camera attached to the Cerro
Tololo Inter-American Observatory (CTIO) 1-m telescope, operated by the SMARTS 
consortium\footnote{\tt http://http://www.astro.yale.edu/smarts} during an observation
run in Nov-Dec 2005. 
This camera is equipped with an STA
4064$\times$4064 CCD\footnote{\texttt{http://www.astronomy.ohio-state.edu/Y4KCam/ detector.html}}
with 15-$\mu$m pixels, yielding a scale of 0.289$^{\prime\prime}$/pixel
and a field-of-view (FOV) of $20^{\prime} \times 20^{\prime}$ at the
Cassegrain focus of the CTIO 1-m telescope. The CCD was operated without binning, at a nominal
gain of 1.44 e$^-$/ADU, implying a readout noise of 7~e$^-$ per quadrant (this detector is read
by means of four different amplifiers).
\\
\noindent

In Table~2 we present the log of our \emph{UBVI} observations, while
in Fig~1  we show astrometrized images, resulting from montaging all the available exposures and filters.
All observations were carried out in
photometric, good-seeing (always less than 1.2 arcsec), conditions. 
Our \emph{UBVI} instrumental photometric system was defined
by the use of a standard broad-band Kitt Peak \emph{UBVI$_{kc}$} set of filters.\footnote{\texttt{http://www.astronomy.ohio-state.edu/Y4KCam/ filters.html}}
To determine the transformation from our instrumental system to the standard Johnson-Kron-Cousins
system, and to correct for extinction, we observed stars in Landolt's areas 
Rubin~149, SA~95, TPhe, PG~0231(Landolt 1992)
multiple times and with different
air-masses ranging from $\sim1.03$ to $\sim2.0$, and covering quite a large color range 
-0.3 $\leq (B-V) \leq$ 1.7 mag.

\begin{table}
\tabcolsep 0.1truecm
\caption{$UBVI$ photometric observations of star clusters.}
\begin{tabular}{lcccc}
\hline
\noalign{\smallskip}
Target& Date & Filter & Exposure (sec) & airmass\\
\noalign{\smallskip}
\hline
\noalign{\smallskip}
Haffner~15   & 2005 Nov 30     & \textit{U} & 60, 1000             &1.00$-$1.02\\
             &                 & \textit{B} & 30,  800             &1.00$-$1.01\\
             &                 & \textit{V} & 15,  400             &1.00$-$1.01\\
             &                 & \textit{I} & 15,  400             &1.00$-$1.01\\
Haffner~4    & 2005 Dec 01     & \textit{U} & 30, 1200             &1.00$-$1.03\\
             &                 & \textit{B} & 20,  900             &1.00$-$1.06\\
             &                 & \textsl{V} & 15,  600             &1.00$-$1.05\\
             &                 & \textsl{I} & 15,  600             &1.00$-$1.04\\
Haffner~11   & 2005 Dec 01     & \textit{U} & 30, 1200             &1.00$-$1.05\\
             &                 & \textit{B} & 20,  900             &1.00$-$1.05\\
             &                 & \textit{V} & 15,  600             &1.00$-$1.05\\
             &                 & \textit{I} & 15,  600             &1.00$-$1.03\\
Berkeley~76  & 2005 Dec 02     & \textit{U} & 30, 1200             &1.05$-$1.08\\
             &                 & \textit{B} & 20,  900             &1.06$-$1.09\\
             &                 & \textit{V} & 15,  600             &1.07$-$1.08\\
             &                 & \textit{I} & 15,   60             &1.06$-$1.09\\
Ruprecht~10  & 2005 Dec 03     & \textit{U} & 30, 1200             &1.05$-$1.14\\
             &                 & \textit{B} & 20,  720             &1.06$-$1.17\\
             &                 & \textit{V} & 10,   30, 540        &1.07$-$1.15\\
             &                 & \textit{I} & 10,  540             &1.06$-$1.14\\
Haffner~7    & 2005 Dec 03     & \textit{U} & 30, 1200             &1.00$-$1.02\\
             &                 & \textit{B} & 20,  900             &1.00$-$1.02\\
             &                 & \textit{V} & 15,  600             &1.00$-$1.01\\
             &                 & \textit{I} & 15,  300             &1.00$-$1.01\\
\noalign{\smallskip}
\hline
\end{tabular}
\end{table}

\subsection{Photometric reductions}

Basic calibration of the CCD frames was done using IRAF\footnote{IRAF is distributed
by the National Optical Astronomy Observatory, which is operated by the Association
of Universities for Research in Astronomy, Inc., under cooperative agreement with
the National Science Foundation.} package CCDRED. For this purpose, zero exposure
frames and twilight sky flats were taken every night.  Photometry was then performed
using the  DAOPHOT/ALLSTAR stand-alone packages. Instrumental magnitudes were extracted
following the point-spread function (PSF) method (Stetson 1987). A quadratic, spatially
variable, master PSF (PENNY function) was adopted, because of the large field
of view of the two detectors. Aperture corrections were then determined in each image
by performing  aperture photometry of a suitable number (typically 10 to 20) of bright, isolated,
stars in the field. Five different apertures were used, going from the small aperture used
for clusters' frame reduction (4-7 pixels, depending on the cluster) 
to the large one used for standard stars (14 pixels). 
A growth curve was then built up to estimate the correction.
These total corrections were found to vary from 0.160 to 0.290 mag, depending
on the filter. The PSF photometry was finally aperture-corrected, filter by filter.

\subsection{Photometric calibration}
During the same observing run we observed the old open cluster Auner~1 (Carraro et al. 2007b), 
the dwarf planet Eris (Carraro et al. 2006)
and three Galactic fields centered on the Canis Major overdensity (Carraro et al. 2008). In these papers
we reported the photometric calibration procedure and the various color equations, which we are not
going to repeat here. The reader is referred to those papers for all the details.
We only remind the reader that typical total errors combining ALLSTAR errors and
calibration errors (see Patat \& Carraro 2001, Appendix A1, for details) are shown in Fig~2.

\begin{figure*}
\includegraphics[width=0.45\hsize]{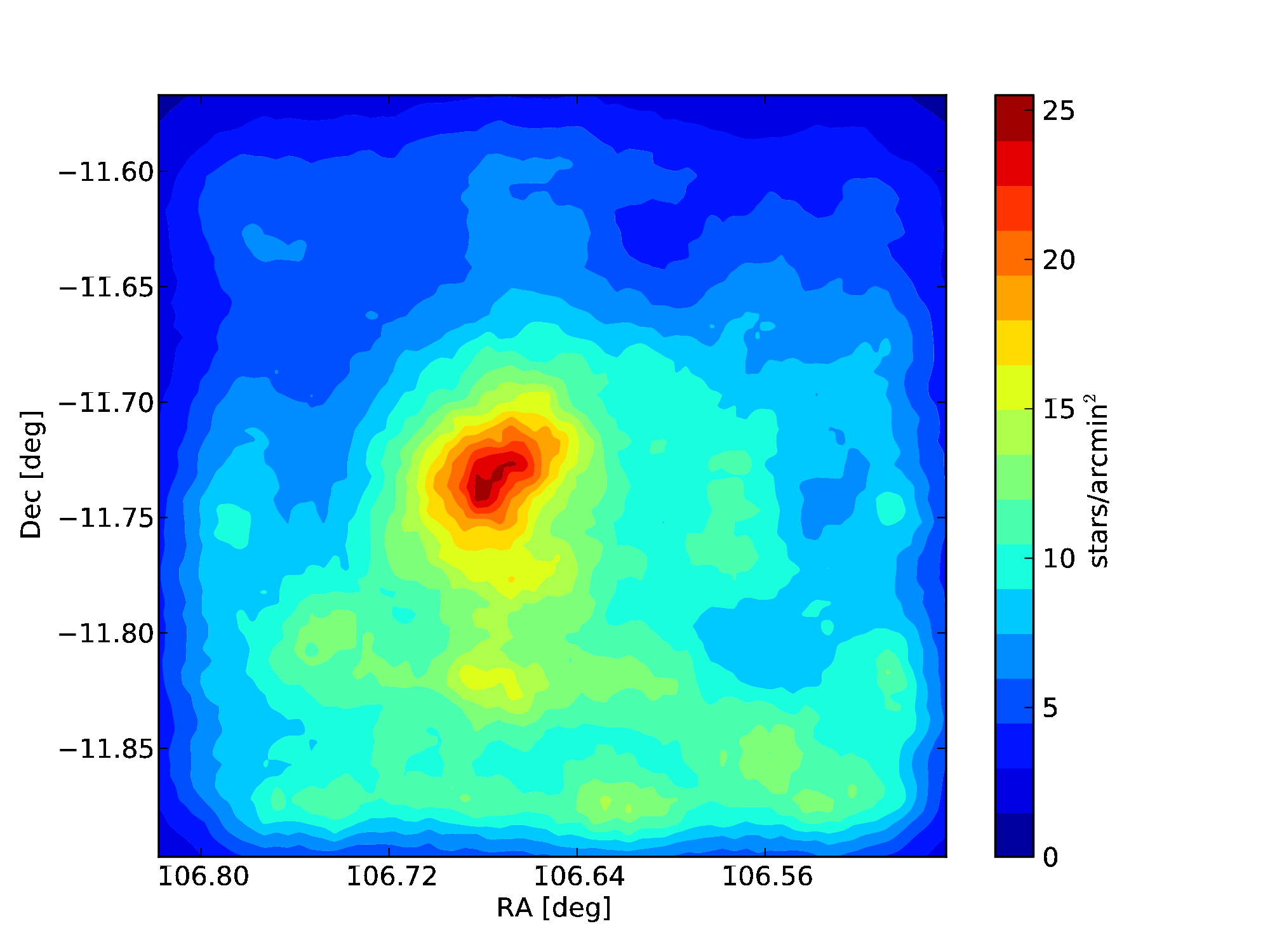}
\includegraphics[width=0.45\hsize]{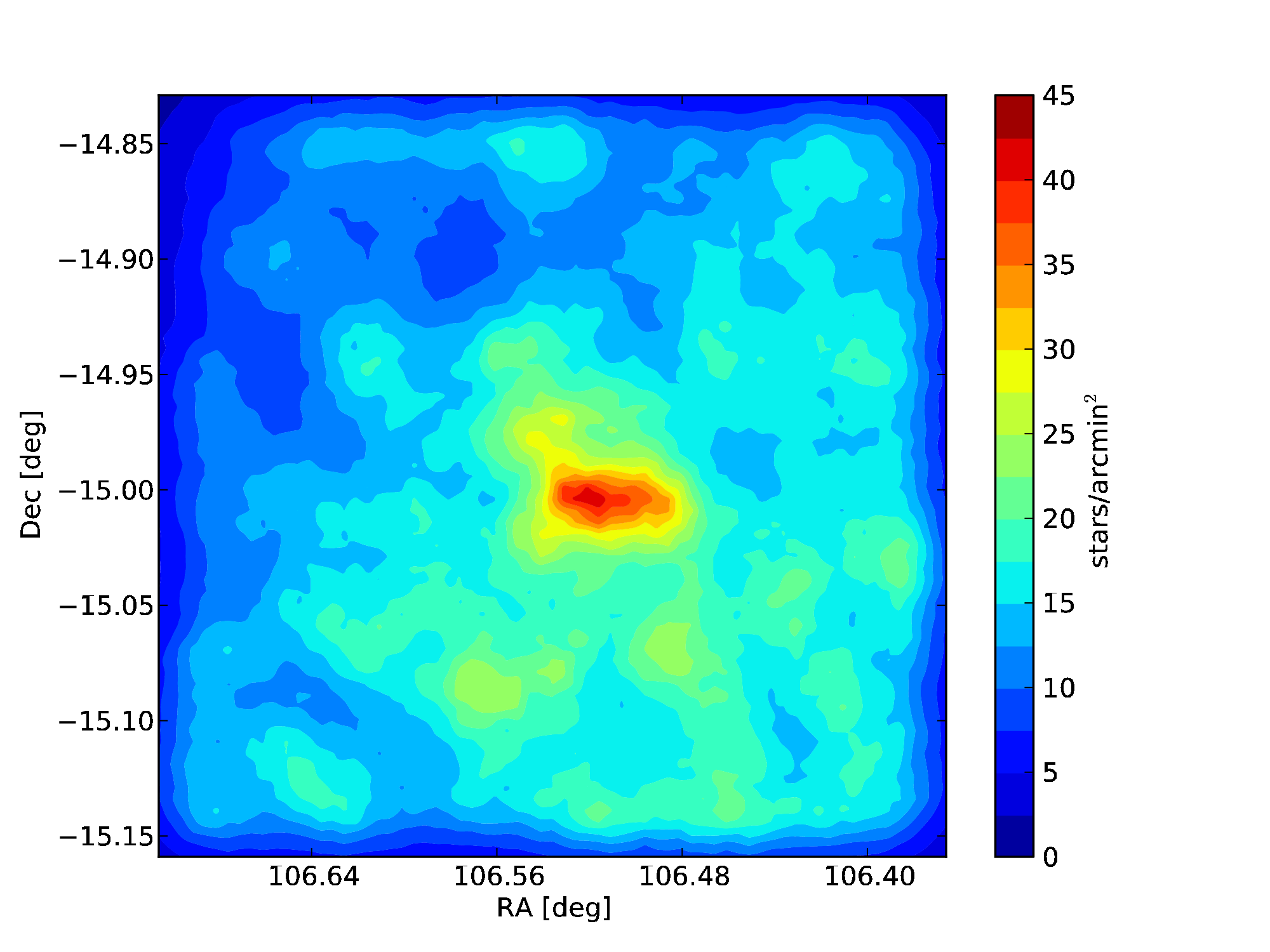}
\includegraphics[width=0.45\hsize]{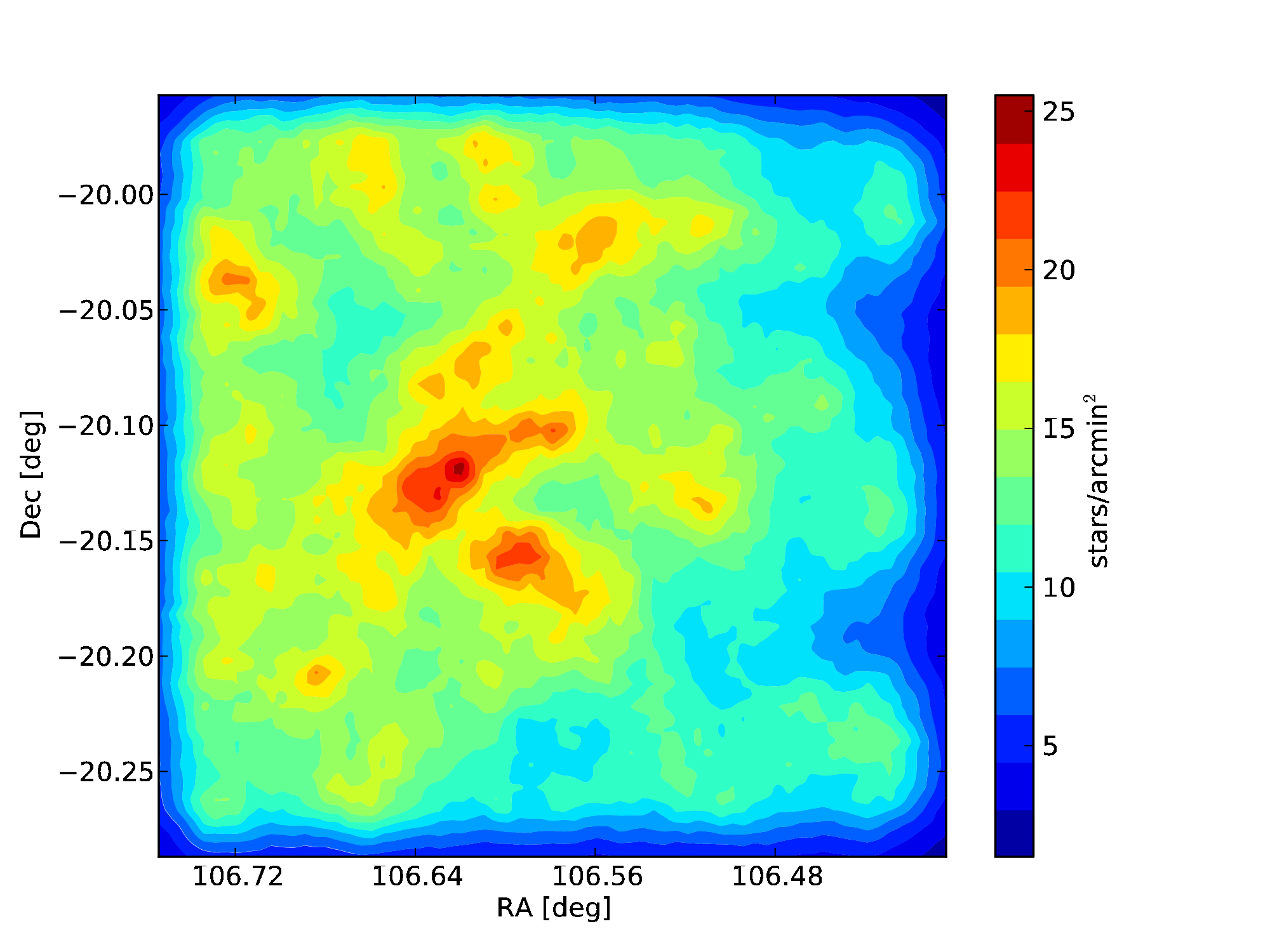}
\includegraphics[width=0.45\hsize]{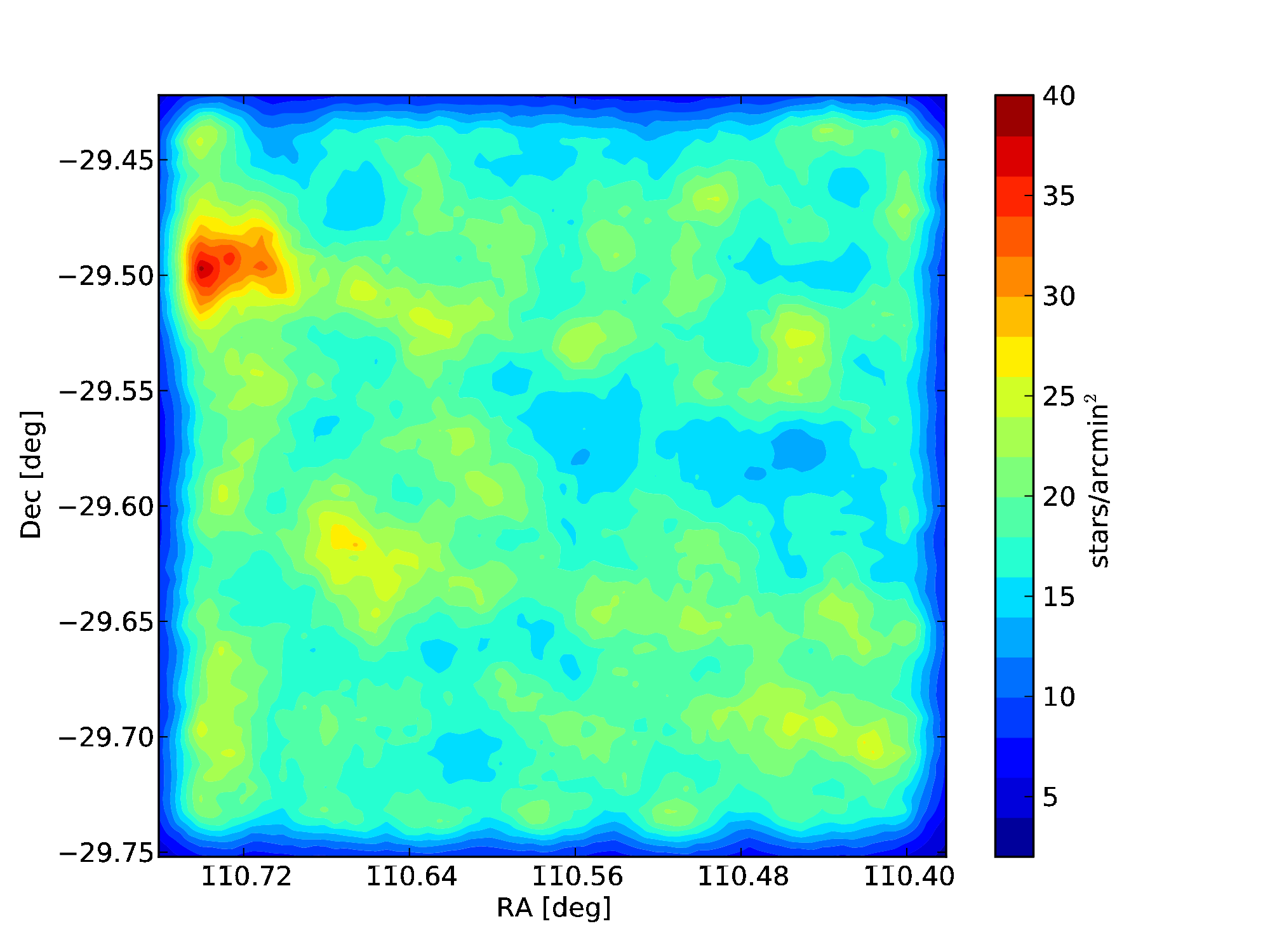}
\includegraphics[width=0.45\hsize]{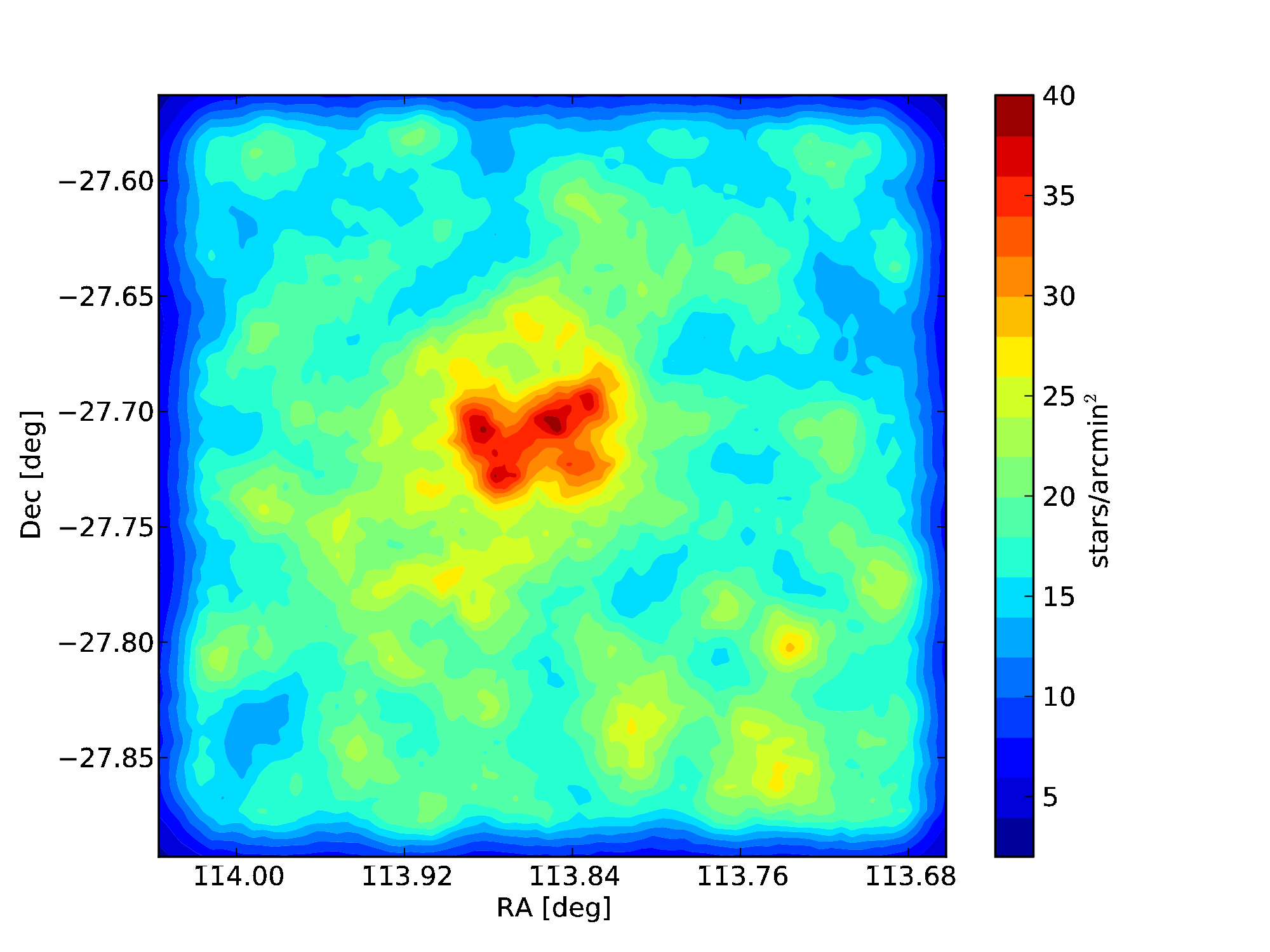}
\includegraphics[width=0.45\hsize]{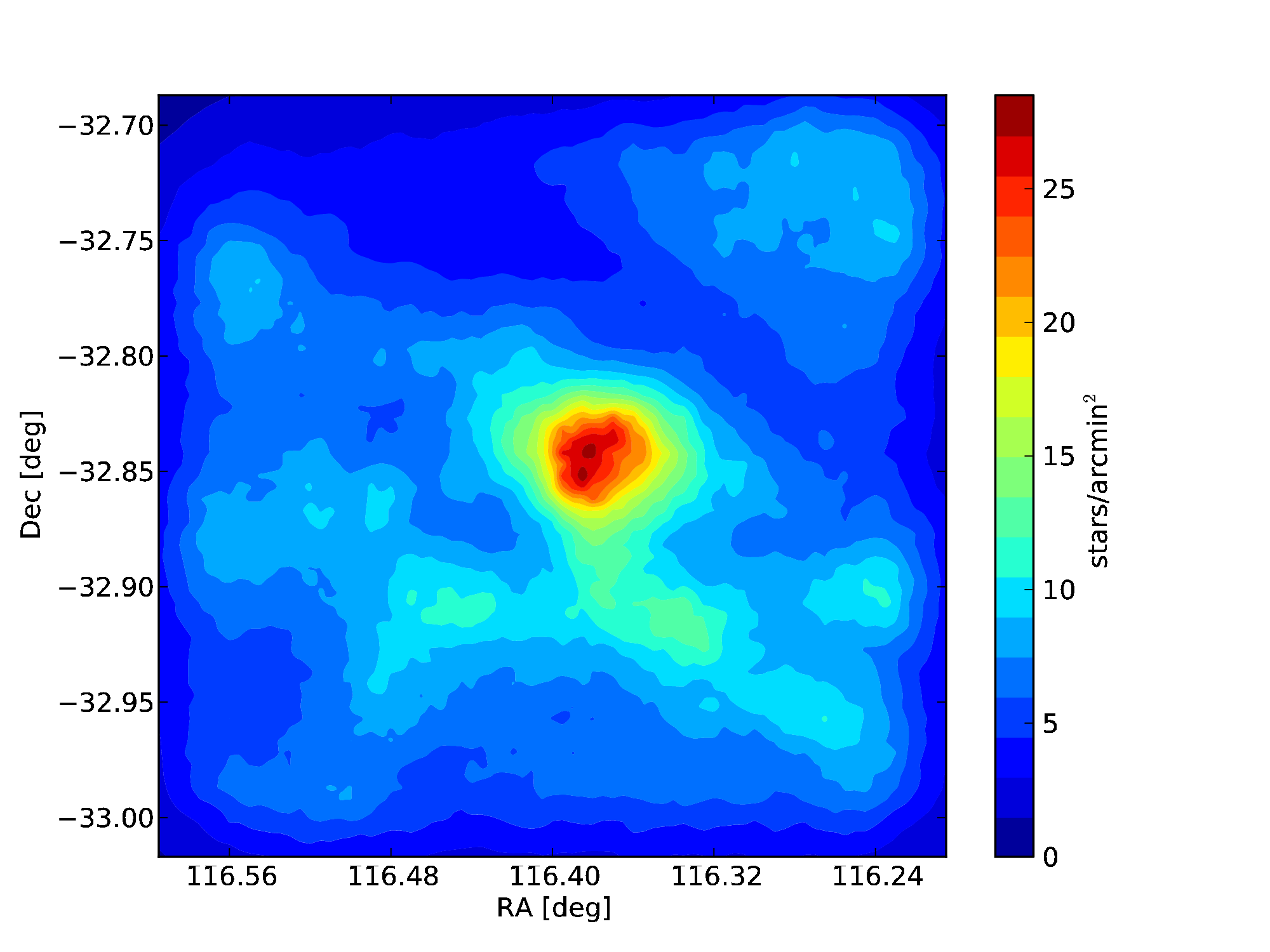}
\caption{ Contour density maps for the 6 clusters under study.
{\it Top raw} (from left to right)-   Berkeley~76, Haffner~4;
{\it middle raw} (from left to right) -  Ruprecht~10,
Haffner~7; {\it bottom raw}(from left to right) -  Haffner~11, Haffner~15.
}
\end{figure*}

\subsection{Completeness and astrometry}

Completeness corrections were determined by running artificial star experiments
on the data. Basically, we created several artificial images by adding artificial stars
to the original frames. These stars were added at random positions, and had the same
colour and luminosity distribution of the true sample. To avoid generating overcrowding,
in each experiment we added up to 20\% of the original number of stars. Depending on
the frame, between 1000 and 5000 stars were added. In this way we have estimated that the
completeness level of our photometry is better than 90\% down to V and I  = 20.5.\\

Each optical catalog was then cross-correlated with 2MASS, which resulted in a final catalog
including \textit{UBVI} and \textit{JHK$_{s}$} magnitudes. As a by-product, 
pixel (i.e., detector) coordinates
were converted to RA and DEC for J2000.0 equinox, thus providing 2MASS-based astrometry, useful
for {\it e.g.} spectroscopic follow-up. The rms of the
residuals in the positions were ∼0.$^{\prime\prime}$17, which is about the astrometric
precision of the 2MASS catalogue (∼0.$^{\prime\prime}$12, Skrutskie et al. 2006 ).\\

All the data discussed in this paper will be made availabe at the WEBDA\footnote{http://www.univie.ac.at/webda/navigation.html} database
maintained by E. Paunzen in Vienna and at VizierR. A sample of the data is shown in Table~3 for the case of Haffner~4.

\renewcommand{\thetable}{3}
\begin{table*}
\caption{A sample of the available photometric data for the star cluster Haffner 4. 99.9999 values are for not available measures.} 
\begin{tabular}{rrrrrrrrrrr}
\hline

ID &  RA & Dec & U & $\sigma_U$& B &  $\sigma_B$ & V &  $\sigma_V$ & I & $\sigma_I$ \\
   & deg & deg & (mag) & (mag) & (mag) & (mag) & (mag) & (mag) & (mag) & (mag)\\
\hline
   1&  106.3883529& -14.8720246&    9.4054&    0.0173&    9.9289&    0.1974&    7.4300&    0.0318&    6.2286&    0.0555\\
   2&  106.5810012& -14.9762229&    9.2274&    0.0154&    9.3976&    0.0157&    9.2511&    0.0161&    9.0499&    0.0516\\
   3&  106.4797390& -15.0037127&   10.5342&    0.0124&   99.9439&   99.9999&   99.9999&   99.9999&   99.9999&   99.9999\\
   4&  106.4823592& -14.9494247&   10.4104&    0.0119&   10.5902&    0.0112&   10.5082&    0.0114&   10.3128&    0.0510\\
   5&  106.4901779& -14.8493842&   99.9999&   99.9999&   99.9999&   99.9999&   99.9999&   99.9990&    8.8061&    0.0534\\
   6&  106.4469765& -14.9727220&   11.2471&    0.0136&   11.2699&    0.0120&   10.7533&    0.0119&   10.0105&    0.0512\\
   7&  106.5754512& -15.0452087&   11.5660&    0.0125&   11.4621&    0.0123&   11.1297&    0.0115&   10.6893&    0.0514\\
   8&  106.6732766& -14.8446577&   12.4649&    0.0218&   13.3356&    0.2078&   10.2985&    0.0329&    9.1951&    0.0673\\
   9&  106.5506869& -14.9631866&   11.8655&    0.0123&   11.8413&    0.0121&   11.3144&    0.0118&   10.6159&    0.0510\\
  10&  106.5745932& -14.9484442&   12.0282&    0.0144&   11.9441&    0.0129&   11.3918&    0.0129&   10.6919&    0.0510\\
\hline
\end{tabular}
\end{table*}

\section{Literature material}
We observed all these clusters back in 2005 and at that time  no studies existed in  literature
on them.\\
However, in the following years, and before the analysis presented in this paper, various studies appeared.
We summarize here the basic results of these investigations. \\

\noindent
Hasegawa et al. (2008) presented CCD BVI photometry of Berkeley~76 and Haffner~4. Their photometry is
typically 3-4 magnitudes shallower than the present study and rarely reaches V $\sim$ 19 mag.
They found (see their Table~4) for Berkeley~76 an age of 1.6 Gyr for Z=0.000 metallicity, a reddening E(V-I)=0.70
and a corrected distance modulus (m-M)$_0$ = 14.39. As for Haffner~4, they found an age of 1.3 Gyr for a metallicity
value of Z=0.008, a reddening E(V-I) = 0.40 and a corrected distance modulus (m-M)$_0$ = 13.24.\\

\noindent
Using the 2MASS catalog, Bica \& Bonatto (2005) derived estimates of the basic parameters of Haffner~11. They
suggest the cluster is 0.95 Gyr old, with a reddening E(B-V) = 0.36 and a corrected distance modulus (m-M)$_0$ = 13.60.
This same cluster was later studied by Piatti et al. (2009) using photometry in the Washington system. These authors find
significantly different results from Bica \& Bonatto (2005), proposing that Haffner~11 is 0.5 Gyr old,
with a reddening E(B-V)= 0.57 and a corrected distance modulus (m-M)$_0$ = 13.90.\\

\noindent
Finally, Haffner~15 was targeted by Paunzen et al. (2006) in their search for peculiar stars in open clusters. In their
study they report an age of 15 Myr, a reddening E(B-V)=1.1 and a corrected distance modulus (m-M)$_0$ = 11.70.\\

\noindent
As for the other two clusters (Ruprecht~10 and Haffner~7), no studies exist in the literature
to our knowledge. 

\begin{figure}
\includegraphics[width=\columnwidth]{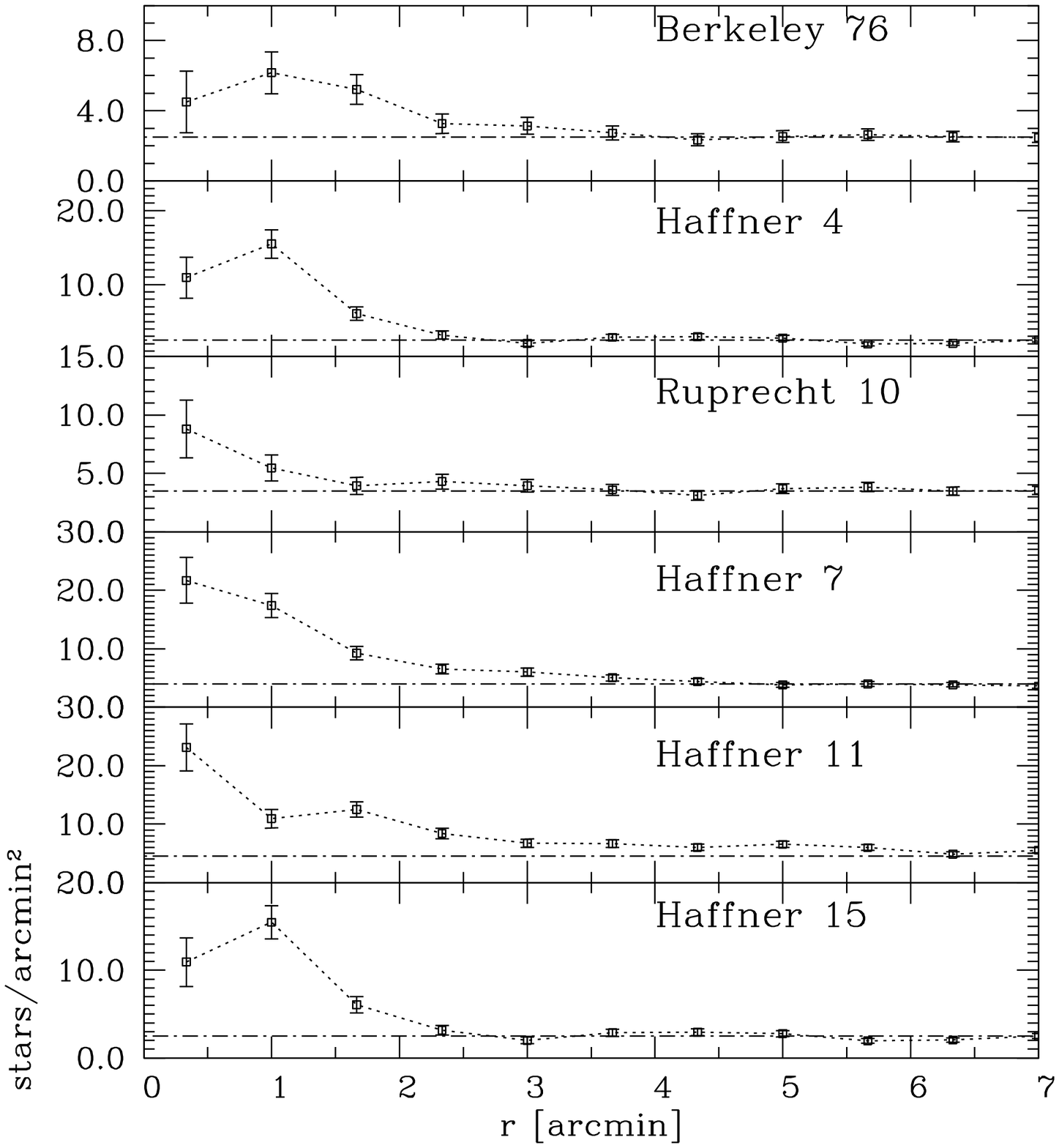}
\caption{Radial density profiles for the 6 clusters under study.}
\end{figure}

 \begin{figure}
   \centering
   \includegraphics[width=\columnwidth]{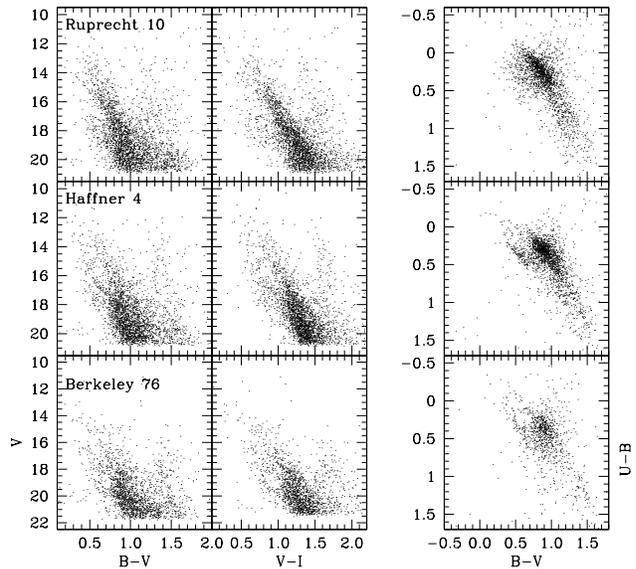}
   \caption{Color-Magnitude and Color-Color Diagrams for Berkeley~76 (lower row), Haffner~4 (mid row) and Ruprecht~10
       (upper row). Only stars having error in the all the pass-bands lower than 0.05 mag are shown.}
    \end{figure}

  \begin{figure}
   \centering
   \includegraphics[width=\columnwidth]{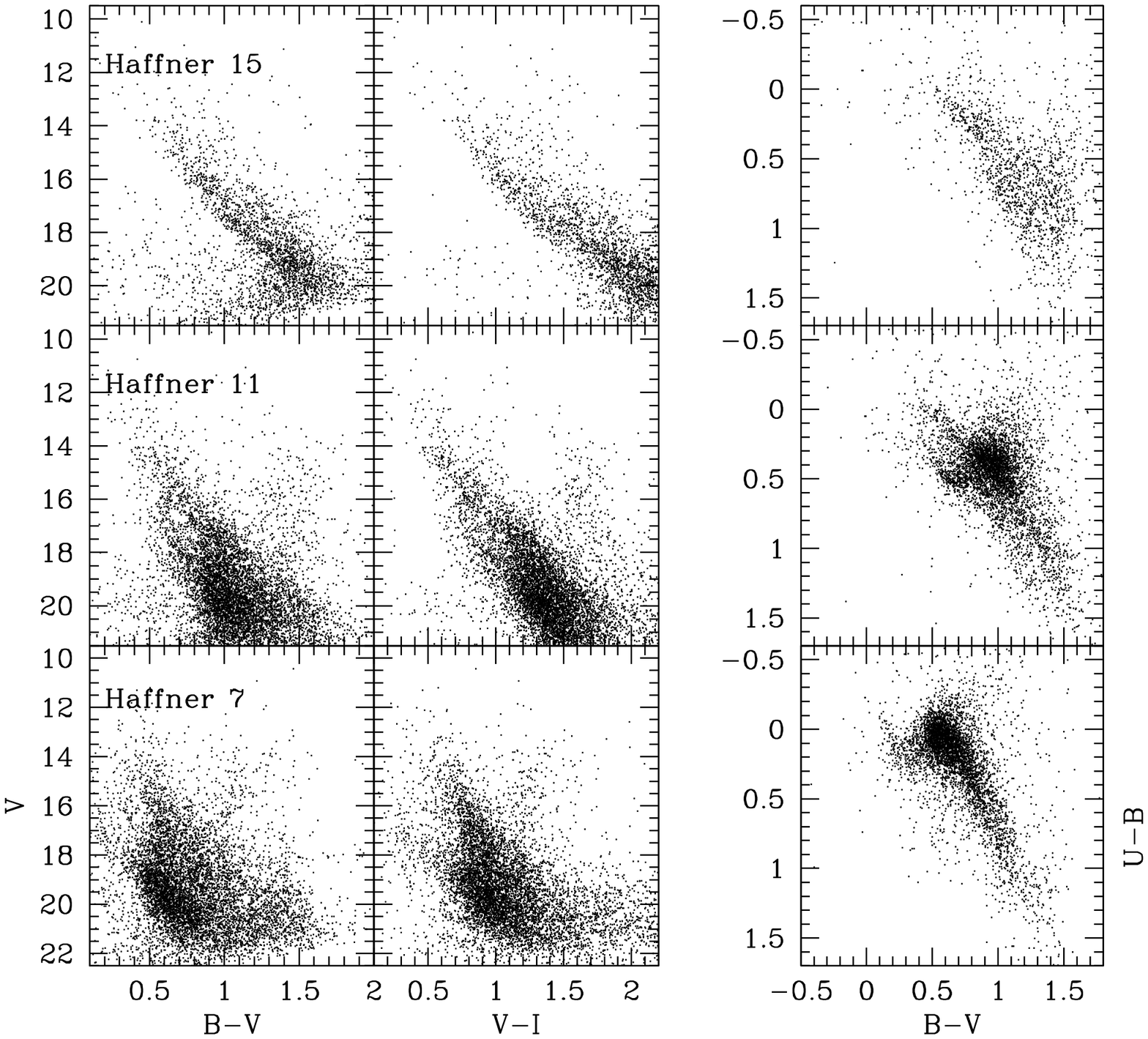}
   \caption{Color-Magnitude and Color-Color Diagrams for Haffner~7 (lower row), Haffner~11 (mid row) and Haffner~15
       (upper row). Only stars having error in all the pass-band lower than 0.05 mag are shown.}
    \end{figure}

\section{Cluster existence and nature}
Before entering into the details of the data analysis, a word of caution is in order. We are here
dealing with poor stellar groups, as can be seen in the CCD images in Fig.~1.
Following Platais et al. (1998), we 
define as a physical group or a gravitationally bound system (at odd with a random sample
of field stars), an ensemble of stars which (1) occupy a limited volume of space, (2)
individually share a common space velocity,  and (3) individually share the same age and
chemical composition, producing distinctive sequence(s) in the Hertzsprung-Russel (H-R) diagram.\\

\noindent
Unfortunately, we do not have kinematic information for these objects, and therefore we are going
to rely on the H-R diagram and star density profiles for deciding on the nature and existence of these
six stellar systems.

\noindent
We start with CCD images to see what the clusters look like. By inspecting  
Fig~1 closely, we notice that all of them stand above the field, and 
fall close to the  center of the Y4KCAM mosaic using catalogued coordinates
(Dias et al. 2002). The only exception is Haffner~7 (mid-right image in Fig.~1)
which we were forced to position in the upper-left corner of the detector to avoid  the V = 2.40 mag $\eta$ Canis Major variable star,
which happens to fall less than 10$^{\prime}$  from Haffner~7 center.
All the clusters exhibit low density against the field, which seems to indicated they are low mass, sparse objects on
the verge of dissolving into the general Galactic field.\\

  \begin{figure*}
   \centering
   \includegraphics[]{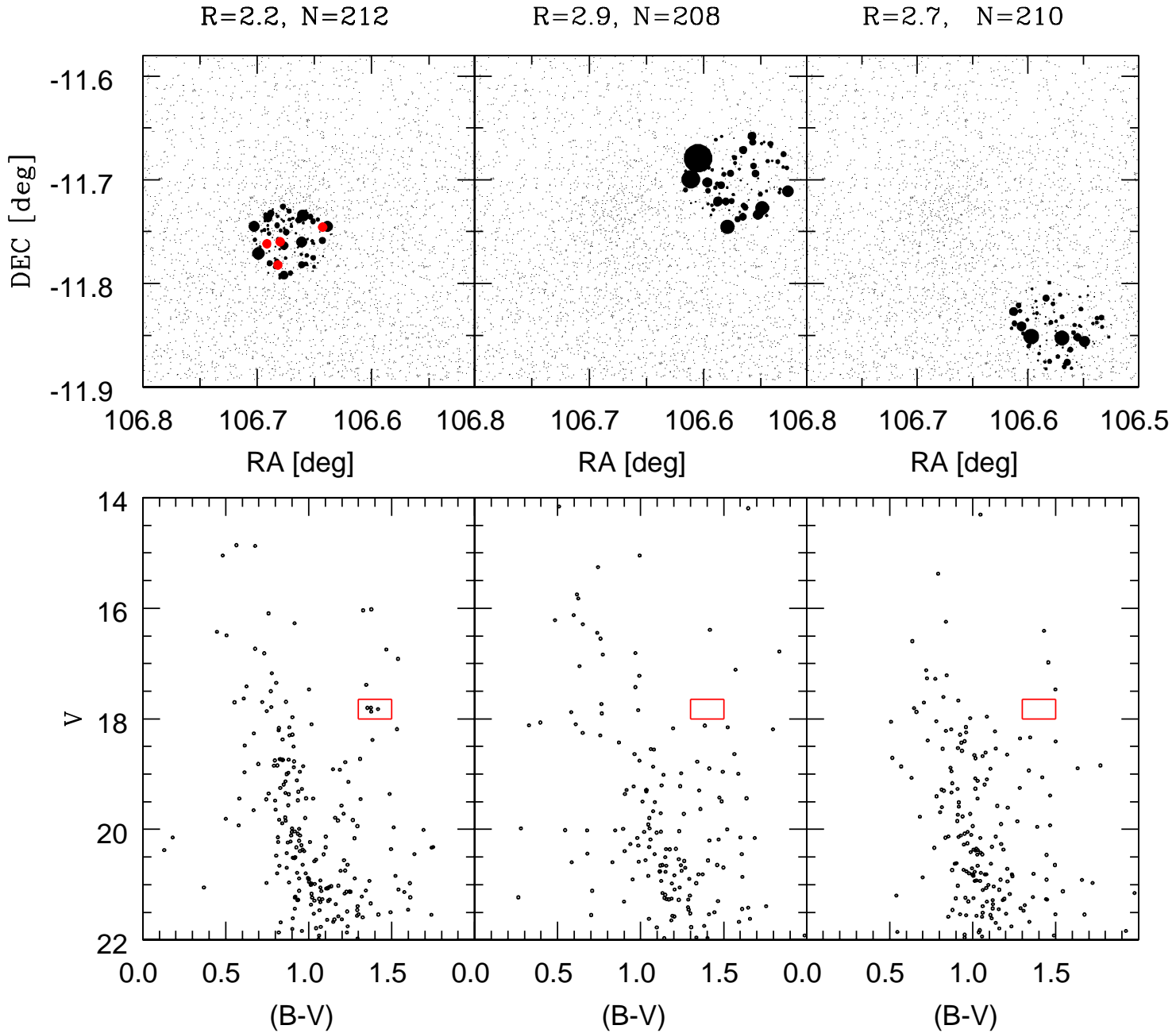}
   \caption{CMDs of Berkeley~76 and surrounding fields. Stars inside these regions are plotted with symbols proportional
   to their magnitude.The red box indicates the position of clump stars. These stars are plotted as red filled circles in the corresponding map.}
    \end{figure*}

  \begin{figure*}
   \centering
   \includegraphics[]{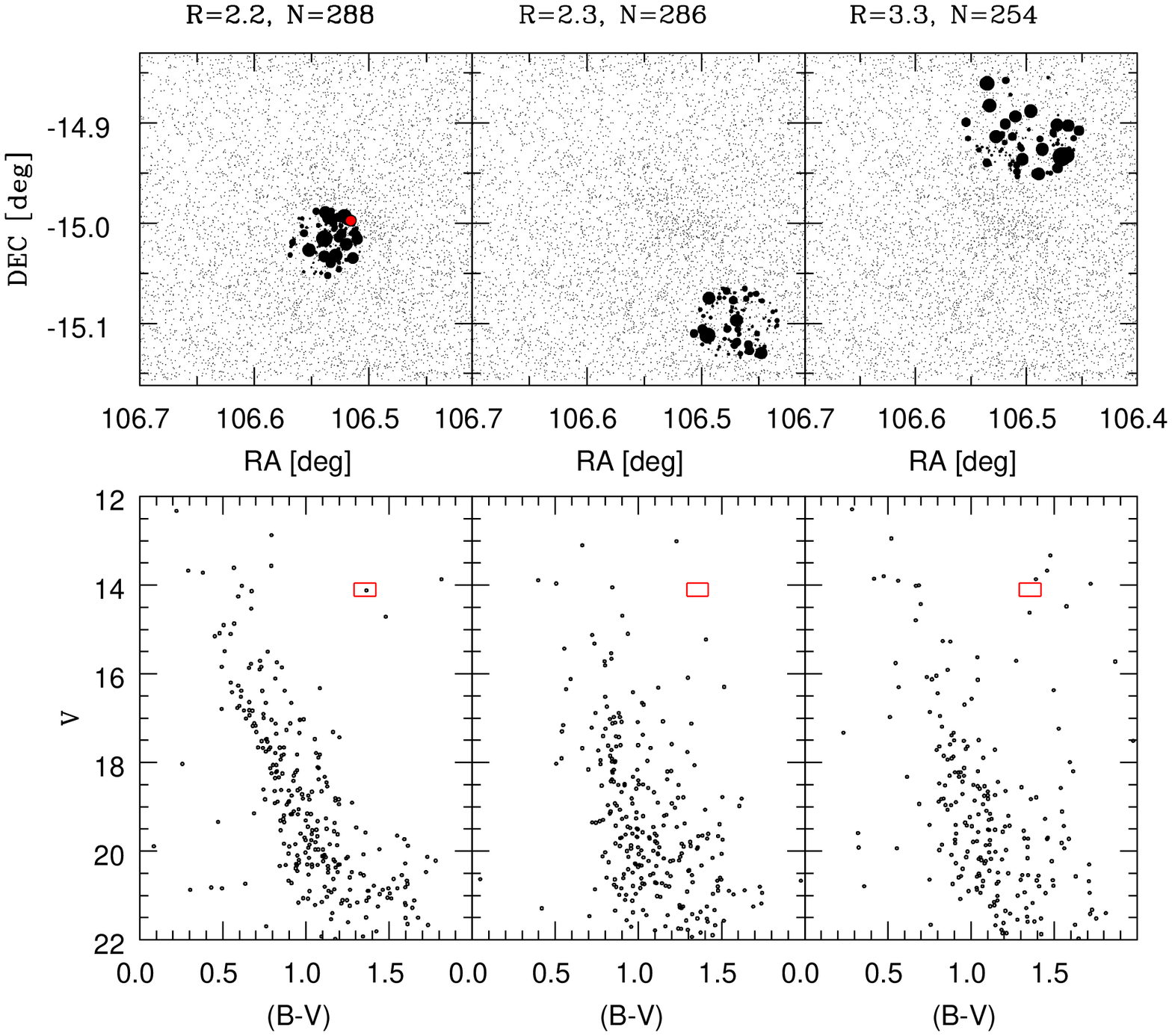}
   \caption{CMDs of Haffner 4 and surrounding fields.Stars inside these regions are plotted with symbols proportional
   to their magnitude.The red box indicates the position of clump stars. These stars are plotted as red filled circles in the corresponding map}
    \end{figure*}

\noindent
We performed star counts on the photometric datasets to derive an estimate of clusters' radial extents.
These have been calculated following the same procedure described in Seleznev et al. (2010) and Carraro \& Costa (2007).\\
Star counts and density contour maps (see Figs.~3 and 4)
confirm the visual impression of CCD images,  and indeed the density of stars in the cluster's area goes from only 2 to 5 times
the density in the surrounding fields, as measured away from the clusters' area (see Janes
\& Hoq 2011 for a similar situation).
Looking at the right columns, where the radial density profiles are shown, one can appreciate how the clusters are in
fact small, with radii of the order of 2-3 arcmin only. The radius has been taken at the  point where star counts
reach the field level, as measured in the field far away from the cluster.
The adopted radii are listed in Table~4, together with
the basic parameters inferred in the next Sects. They are in agreement with the visual estimates
reported in Dias et al. (2002).\\
\noindent
We therefore conclude that all these clusters are indeed made of groups of stars spatially concentrated and
standing above the general Galactic field.\\

\noindent
Besides, looking at the middle panels of Figs.~3, where contour maps are presented, one can
see how the two-dimensional structure of these clusters is far from spherical.
They show elongated shapes, stretched in one or more directions, and, in some cases, with
more than one density peak in the cluster area. We are tempted to interpret
these complicated structures as evidence that the clusters are undergoing
dissolution due to the interation with the Galactic environment. The ages we find for them,
larger than 1 Gyr (see next Sects), confirm this fact, since the typical lifetime for a Galactic cluster
in the Milky Way is around 200 million years (Pavani \& Bica 2007, Wielen 1971).
The only execption - and a confirmation of this scenario - is the younger cluster Haffner~15, which looks
clearly spherical. We stress, anyway, that these are qualitative conclusions driven by the photometric data only,
and that a dynamical study employing radial velocities
of individual stars can more firmly asess the dynamical status of these poor systems.\\

\section{Color Magnitude Diagrams of clusters and surrounding fields}
In Figs.~4 and 5 we present the Color-Magnitude (CMD) and Color-Color (CCD) diagrams of the regions
containing the 6 clusters
under study. Only stars having photometric errors lower than 0.05 mag. in all the four filters are plotted.
In a wide field as the one we used (20 arcmin on a side) none of the clusters clearly emerges from the field in the CMDs,
except for Haffner~15 (upper row in Fig~6).
All the clusters are hidden inside a rich field star population. The latter shows the typical
features  of any stellar field projected toward the Canis Major overdensity, namely a {\it blue plume} of young stars
and a prominent blue faint sequence which runs to the left of the nearby stars Main Sequence (MS) and crosses
it at 16.5$\leq V \leq$18.5, depending on the field.
These features have been described and extensively interpreted {\it e.g.} in Carraro et al. (2008, 2010),
to which we refer the reader for further details.\\

\noindent
The aim of this paper instead is to derive the fundamental parameters of these 6 clusters, and for this reason we are going to
extract photometry only for those stars which fall inside the cluster radius, as estimated in Section 1.5.
This will have the effect of alleviating the field star contamination, which we expect to be significant since these lines
of sight (see Table~1) are projected toward the warped thin disk (Carraro et al.2008).\\

\noindent
We are not going to present any statistical cleaning of the clusters' CMD, since the number 
statistics in this clusters' sample is poor,
and would generate artificial clumps and voids, which are difficult to be removed, and would 
complicate the interpretation of the diagrams.
In the series of Figures 7 to 12, we present the CMD of each cluster and two realizations of the surrounding field,
with the aim to highlight the cluster against the general field.
This is possible thanks to the small size of the clusters and the wide field of our observations.
In each of Figs. 6 to 11,
the cluster CMD (lower left panel) is constructed by considering all the stars falling inside a circle whose radius is
close to the cluster radius,
as estimated in Sect.~3. The corresponding area in the cluster map is shown in the upper left panel.\\

\noindent
We then constructed CMDs for two different regions outside the cluster border, having
about the same number of stars as the cluster region. 
In each figure the adopted radii are indicated together with the number of stars detected in each zone. 
In all cases the cluster region results to be significantly more populated than the corresponding field
regions, which we needed to enlarge to include a comparable number of stars as in the cluster area.\\

\noindent
  Here below, we provide individual comments  on a cluster by cluster basis.\\
\noindent

  \begin{figure*}
   \centering
   \includegraphics[]{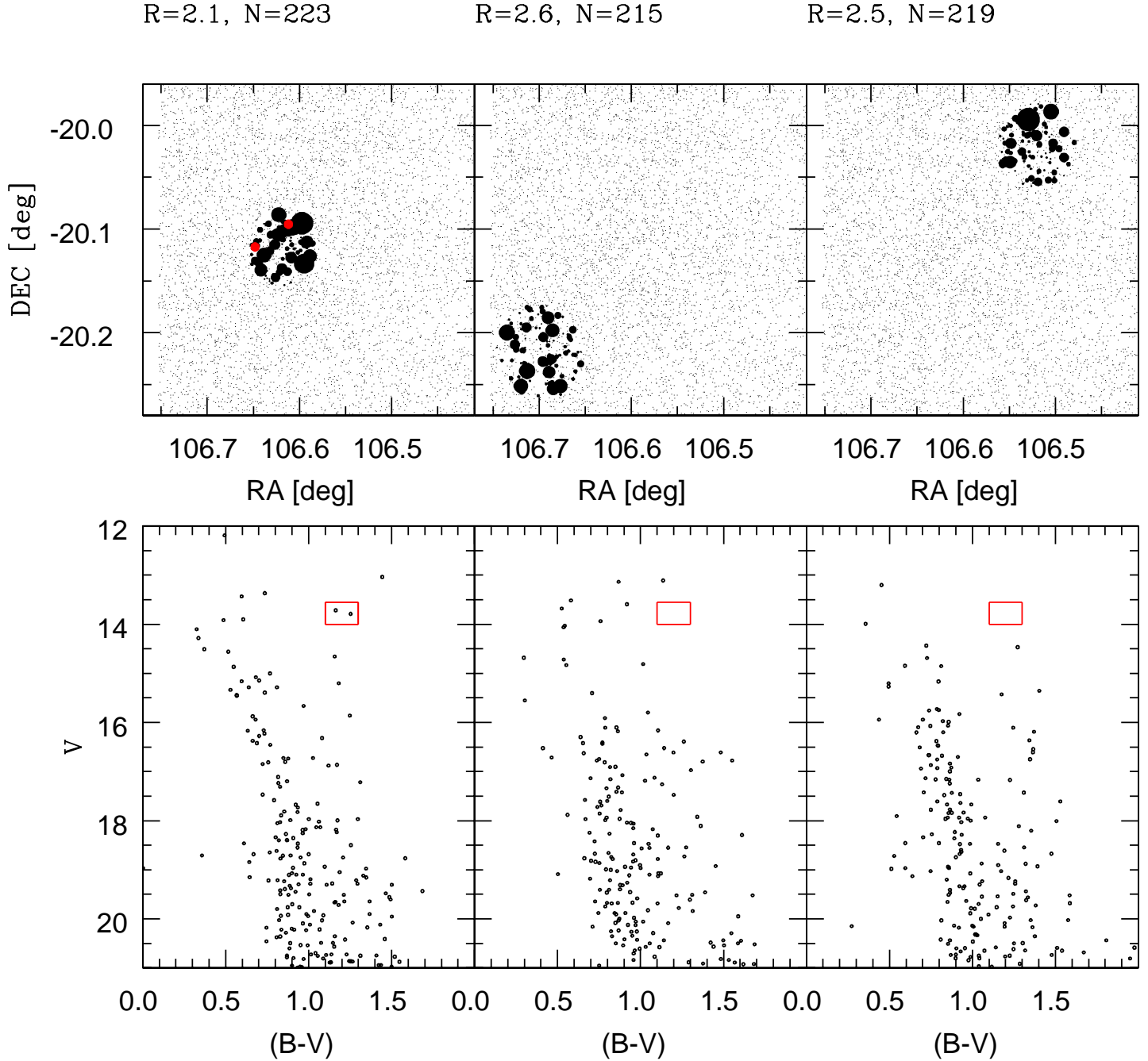}
   \caption{CMDs of Ruprecht 10 and surrounding fields. Stars inside these regions are plotted with symbols proportional
   to their magnitude. The red box indicates the position of clump stars. These stars are plotted as red filled circles in the corresponding map}
    \end{figure*}

  \begin{figure*}
   \centering
   \includegraphics[]{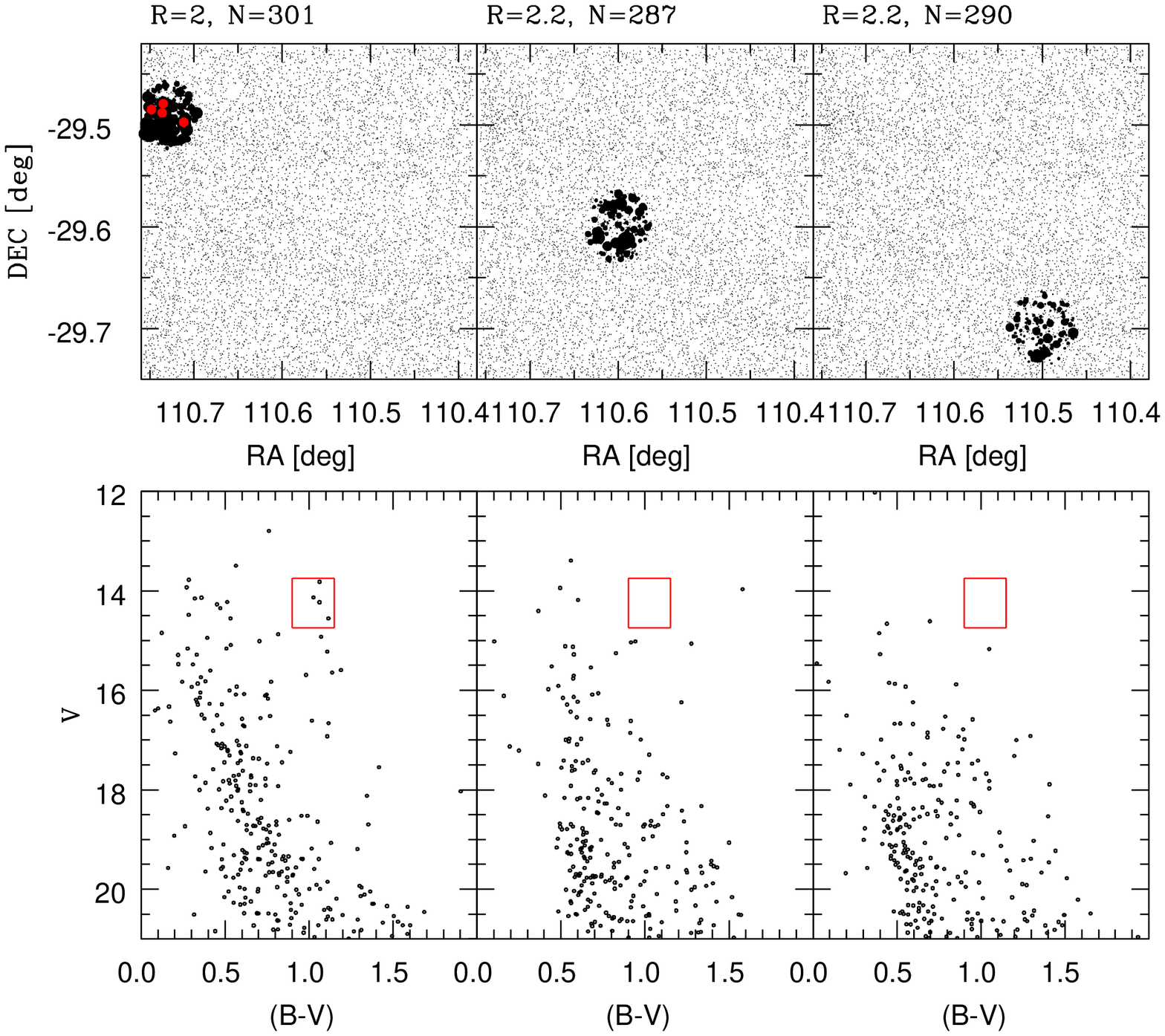}
   \caption{CMDs of Haffner 7 and surrounding fields. Stars inside these regions are plotted with symbols proportional
   to their magnitude. The red box indicates the position of clump stars. These stars are plotted as red filled circles in the corresponding map}
    \end{figure*}

  \begin{figure*}
   \centering
   \includegraphics[]{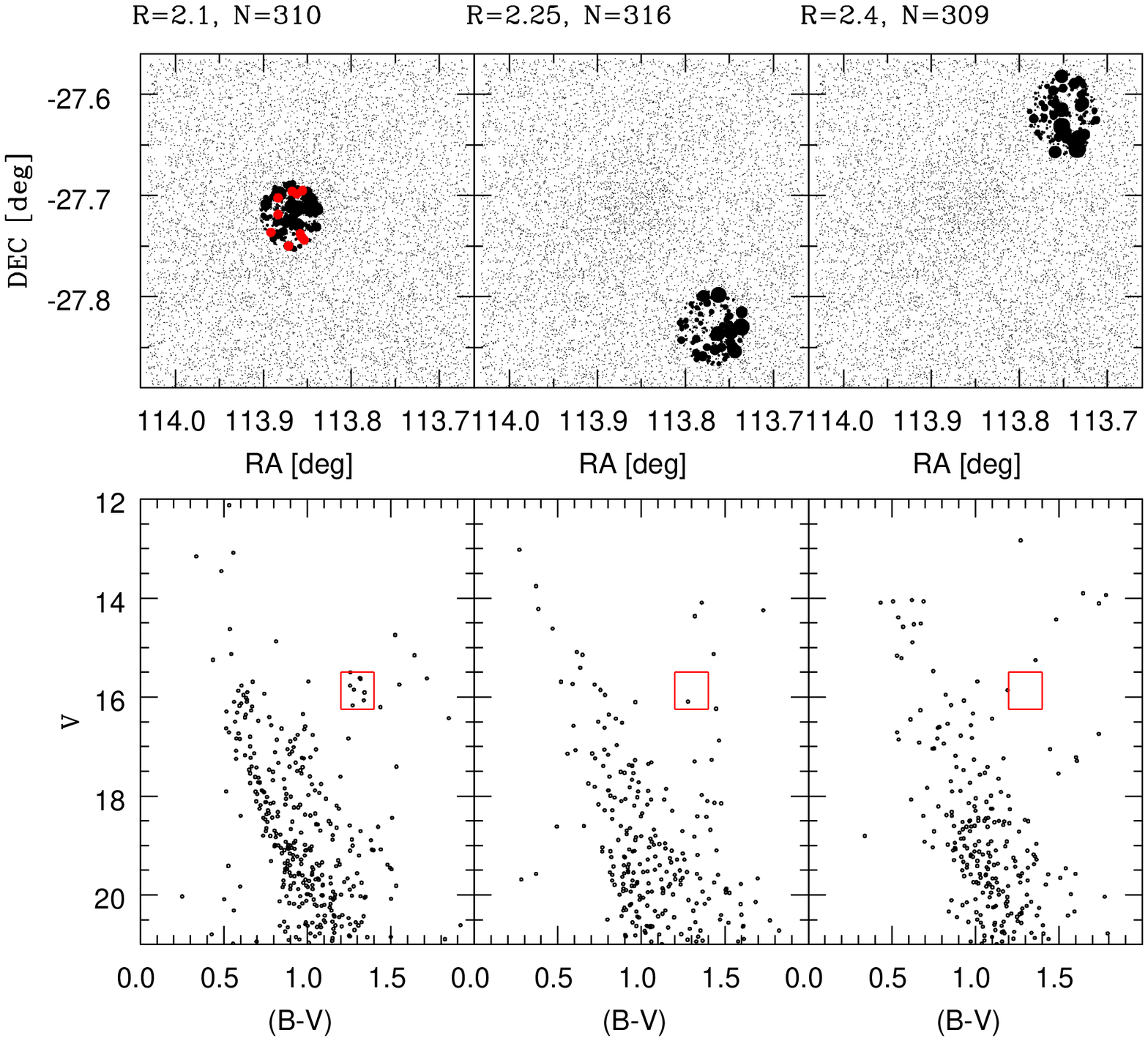}
   \caption{CMDs of Haffner 11 and surrounding fields. Stars inside these regions are plotted with symbols proportional
   to their magnitude. The red box indicates the position of clump stars. These stars are plotted as red filled circles in the corresponding map}
    \end{figure*}

 \begin{figure*}
   \centering
   \includegraphics[]{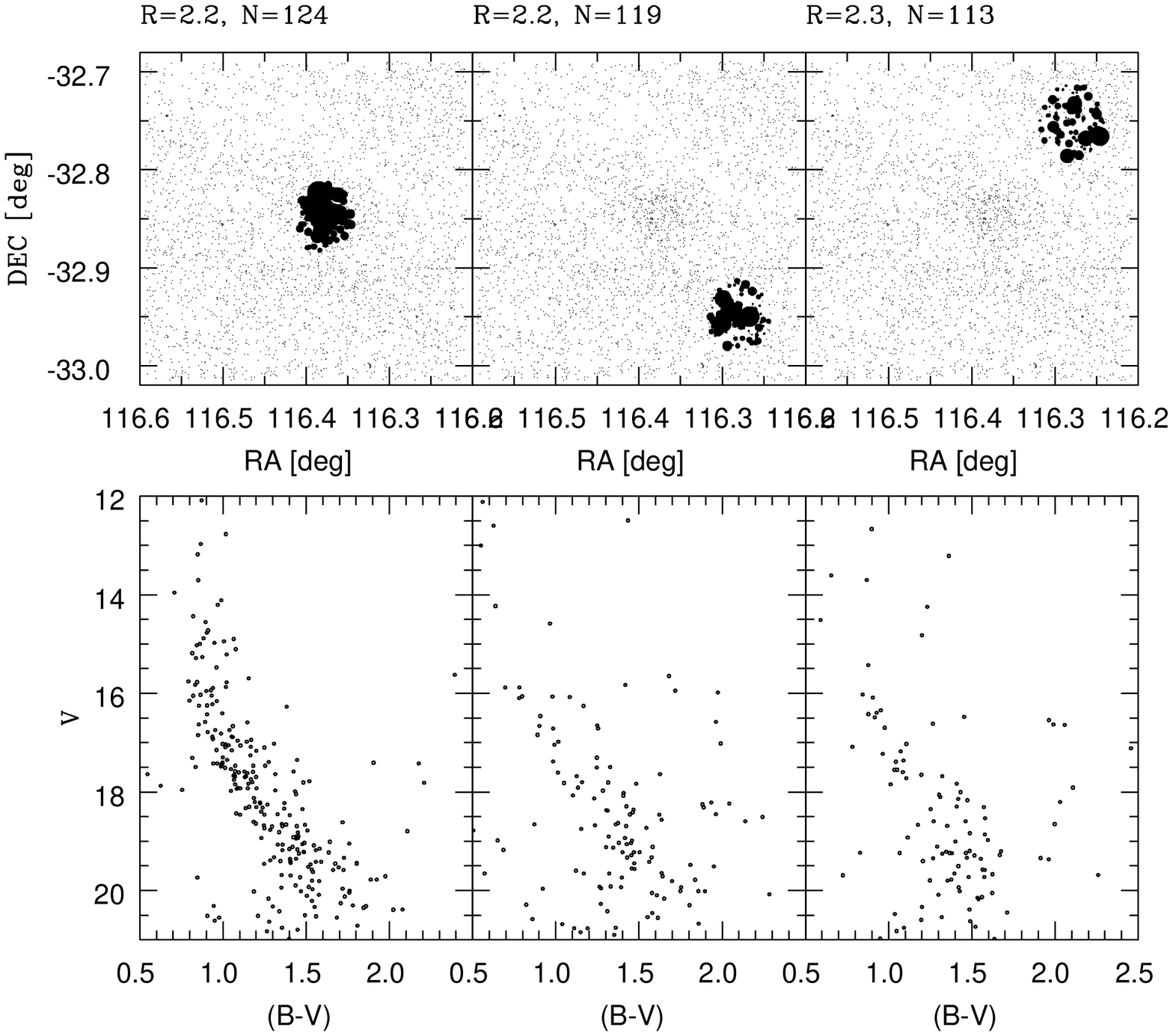}
   \caption{CMDs of Haffner 15 and surrounding fields. Stars inside these regions are plotted with symbols proportional
   to their magnitude.}
    \end{figure*}

\noindent
{\bf Berkeley~76: see Fig~7.}
The CMD in the cluster region (left panel) shows a conspicous main sequence (MS) which
suddendly drops at V $\sim$ 18.8. The bluest point is at V $\sim$ 19.5, which we consider
as the MS turn off point (TO). Another interesting feature is the compact group of 4 stars
at V $\sim$ 17.9, B-V $\sim$ 1.4, which does not have any counterpart in the field,
as indicated by the red box circumscribing it. 
 These stars are plotted as red dots in the corresponding  map. Their position inside the cluster
region lends further support to their identification as red clump stars members to the cluster.
We are going to consider this as the cluster red giant branch (RGB) clump.
We believe  this set of CMDs convincingly shows that we are facing a real star cluster. We shall derive 
estimates of its
basic parameters in the next Section. \\

\noindent
{\bf Haffner~4: see Fig~8.}
Also in this case the cluster CMD reveals a MS significantly more populated than in the field star
CMDs. It is difficult to say where the TO is located, but we tentatively identify it at V $\sim$ 15.0,
B-V $\sim$ 0.5. We do see a possible trace  of a RGB clump in the star that we enclosed
in the red box. Two redder stars can be part of the RGB as well.
 This star is plotted as a red dot in the corresponding  map. Its position inside the cluster
region lends further support to its identification as red clump star member to the cluster.
We shall derive estimates of its
basic parameters in the next Section. \\

\noindent
{\bf Ruprecht~10: see Fig~9.}
The interpretation of Ruprecht~10 CMD is not straitghforward at all.
We tentatively consider as the MS the bright sequence terminating at V $\sim$ 14, B-V $\sim$ 0.3,
while the vertical sequence terminating at V $\sim$ 18.5 is produced by field stars, since
it is visible also in the field star CMDs.
Moreover, we consider as clump stars the 2 stars at V $\sim$ 14, B-V $\sim$ 1.4. 
These stars are plotted as red dots in the corresponding  map. Their position inside the cluster
region lends further support to their identification as red clump stars members to the cluster.
We are going to provide estimates of the fundamental parameters in the next Section.
\\

\noindent
{\bf Haffner~7: see Fig~10.}
This is by far the most complicated object, because we could not cover it completely,
as discussed in Sect.~3. 
The cluster CMD looks different from the field CMDs, and shows a clear MS.
The bluest point is at V $\sim$ 16.50, which we consider
as the MS turn off point (TO). Another interesting feature is the sparse group of 4 stars
at V $\sim$ 14.5, B-V $\sim$ 1.1, which does not have any counterpart in the field,
as indicated by the red box circumscribing it. 
 These stars are plotted as red dots in the corresponding  map. Their position inside the cluster
region lends further support to their identification as red clump stars members to the cluster.
We conclude that in the region of Haffner~7 there is a star concentration,
and this concentration shows distinctive features in the CMD.
We are going to provide estimates of the fundamental parameters in the next Section.\\

\noindent
{\bf Haffner~11: see Fig~11.}
This is undoubtedly  a nice intermediate-age cluster, with a conspicuous clump of stars (enclosed in a red box) at V $\sim$ 16, B-V
$\sim$ 1.3.  These stars are plotted as red dots in the corresponding  map. Their position inside the cluster
region lends further support to their identification as red clump stars members to the cluster.
The MS TO is located at V $\sim$ 16.5. The vertical sequence which drops at V $\sim$ 16.5
redward of the cluster MS is produced by the field, and appears also in the field star CMDs.
Haffner~11 is the most obvious cluster amongst the ones discussed so far. \\

\noindent
{\bf Haffner~15: see Fig~12.}
From the appearence of the cluster area CMD we infer that Haffner 15 is a young cluster, with
a TO at V $\sim$ 15.5 and no indication of RGB stars. The MS is prominent and much more
populated than in the field star CMDs, where the typical vertical sequence from the disk
is visible. The color of the TO indicates that the cluster is significantly reddened.
We are going to provide estimates of the fundamental parameters in the next Section. \\

\section{Derivation of clusters' basic parameters}
The CMDs generated with only the cluster region (as defined in Sect. 1.5)  stars are here compared with isochrones
extracted from the Padova suite of models (Marigo et al. 2008). This allows us to infer estimates
of clusters age, distance and reddening, by assuming conservative values of the metallicity.
We are fully aware that this process is highly subjective, and for this reason we consider the values we obtain as estimates,
awaiting studies which can provide measures of the clusters' metal abundance.\\

\noindent
In fact, we are here facing 
the  well-known problem of associating reliable errors
to distance and age. Without precise estimates of reddening
and metallicity, it is extremely difficult to perform a proper
error assessment. In theory this would imply a full error propagation which
would in general produce
a very  large yper-volume in the parameters' space with many
solutions which would not pass a  simple by-eye inspection.\\

\noindent
We will, therefore, limit ourselves to provide fitting errors
for the cluster reddening and apparent distance moduli, being totally aware
that they most probably are only rough lower limits awaiting improvements as
soon as more precise metallicity measurements will be available.
However, in deriving distance, a full propogation is done
taking into account the whole range of values for reddening and distance modulus.
Finally, as far as the ages is concerned, only fitting errors are reported, adopting
solar metallicity (see below).

\noindent
For consistency with previous studies we shall adopt 8.5 kpc as the Sun distance to the Galactic center,
and 3.1 for the ratio of total to selective absorption $R_V=\frac{A_V}{E_{B-V}}$ , which Moitinho (2001) demonstrated to be of 
general validity in this Milky Way sector.\\

\noindent
{\bf Berkeley~76: see Fig~13}\\
The CMDs for this cluster have been derived plotting all the stars within 2.5 arcmin
from the cluster center. Still the contamination is significant, and makes the exercise
of fitting an isochrone quite challenging.
As previsouly discussed, we consider as cluster Red Giant Branch (RGB) clump the group of 4 stars at V$\sim$ 17.85, (B-V) $\sim$ 1.35.
Such a small number of clump stars is not unusual in old open clusters, as one can see also in
very recent studies (e.g. King~8 or Berkeley~23, Cignoni et al. 2011). The TO is 
located at V $\sim$ 19.5, (B-V) $\sim$ 0.85.

If we use this interpretation of the cluster CMD, 
we end up with an age of about 1.5 Gyr. The over-imposed - half solar (Z=0.008) - isochrone
fits also the slope and shape of the MS. 
The mismatch in color for the clump is of the order of $\Delta (B-V) \sim $ 0.08 mag, and can be
due to a variety of reasons, like color tranformations and/or  mixing length calibration issues (see Carraro \& Costa
2007, and Palmieri et al. 2002). 
We also tried solar metallicity, which does not seem suitable for an outer
disk cluster, and found that for a comparable age, the fit to the MS is acceptable, but
the mismatch in color with the cluster clump is much more severe.\\
For the adopted  age and metallicity, Berkeley~76 has a reddening E(B-V)=0.55$\pm$0.1 (E(V-I)=0.75$\pm$0.10)
and a distance modulus $(m-M)_V = 17.15\pm0.20$. Uncertainties in reddening and distance have been estimated
by eye, shifting the isochrone along the horizontal and vertical directions, respectively, and 
estimating the range of values that yield acceptable fits.
As a consequence, the heliocentric distance is 12.6 kpc, and the distance
from the Galactic center is 17.4 kpc, making this cluster one of the most peripheral old open cluster
in the outer disk (Carraro et al. 2007).\\

\noindent
These results are in basic agreement with Hasegawa et al. (2008), except for the distance. Looking at their results (their Fig~2)
for Berkeley~76, one can indeed notice that the fit to the magnitude of the red clump is offset by almost half a magnitude, with
the isochrone clump being too bright. This explains, in turn, the smaller distance in their study. \\

\noindent
{\bf Haffner~4:  see Fig~14.}\\
As for the previous cluster, we selected only the stars inside the cluster radius (see Table~4).
In this case we use solar metallicity isochrones, since the cluster is presumably 
located closer than Berkeley~76.
Haffner~4 looks scarcely populated, and severely contaminated by field stars.
The best-fit isochrone is for an age of 0.5 Gyr, which nicely follows the shape of the MS and
the TO curvature, at V $\sim$ 15.0, (B-V) $\sim$ 0.5.
No obvious indications of a red clump are present. 
The fit shown in Fig.~7 yields a reddening E(B-V) = 0.50$\pm$0.10 
(E(V-I)= 0.63$\pm$0.10), and a distance
modulus (m-M)$_V$=14.80$\pm$0.15. Once corrected, this implies a distance from the Sun of 4.4 kpc,
and of 11.9 kpc from the Galactic center. \\

\noindent
In this case our results imply an age lower than the one suggested by  Hasegawa et al. (2008),
and a solar metallicity.\\

\noindent
{\bf Ruprecht~10: see Fig~15.}\\
This cluster is sparse and has quite a distorted shape, which 
might indicate a stage of advanced disgregation. 
The MS and the evolved region of the CMD are scarcely populated. The TO
is located at V $\sim$ 15.0, (B-V) $\sim$ 0.5, and the clump
at V $\sim$ 14.0, (B-V) $\sim$ 1.25.
By considering the stars within the cluster estimated radius,
we obtained an acceptable fit for
a half-solar metallicity isochrone of 1.1 Gyr,
as shown in Fig.~8. This fits provides a reddening E(B-V) =0.28$\pm$0.10
and a distance modulus (m-M)$_V$=13.40$\pm$0.10.
This, in turn, yields a heliocentric distance of 2.9 kpc, and a distance
from the Galactic center of 10.5 kpc.\\

\noindent
{\bf Haffner~7: see Fig~16.}\\
The TO
is located at V $\sim$ 16.0, (B-V) $\sim$ 0.70, and the clump
at V $\sim$ 14.5, (B-V) $\sim$ 1.5.
By considering the stars within the cluster estimated radius,
we obtained an acceptable fit for
a half-solar metallicity isochrone of 1.5 Gyr,
as shown in Fig.~8. This fits provides a reddening E(B-V) =0.13$\pm$0.10
and a distance modulus (m-M)$_V$=13.65$\pm$0.10.
This, in turn, yields an heliocentric distance of 4.5 kpc, and a distance
from the Galactic center of 11.3 kpc.\\

  \begin{figure}
   \centering
   \includegraphics[width=\columnwidth]{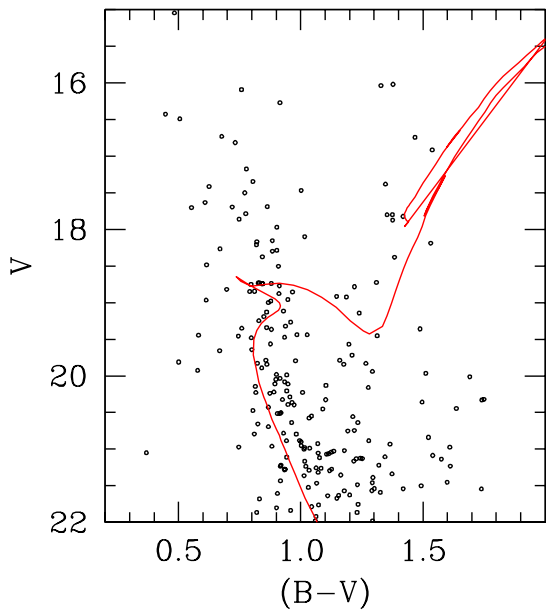}
   \caption{Isochrone solution for Berkeley~76 in the V/B-V CMD. Fitting parameters are summarized in Table~4}
    \end{figure}

  \begin{figure}
   \centering
   \includegraphics[width=\columnwidth]{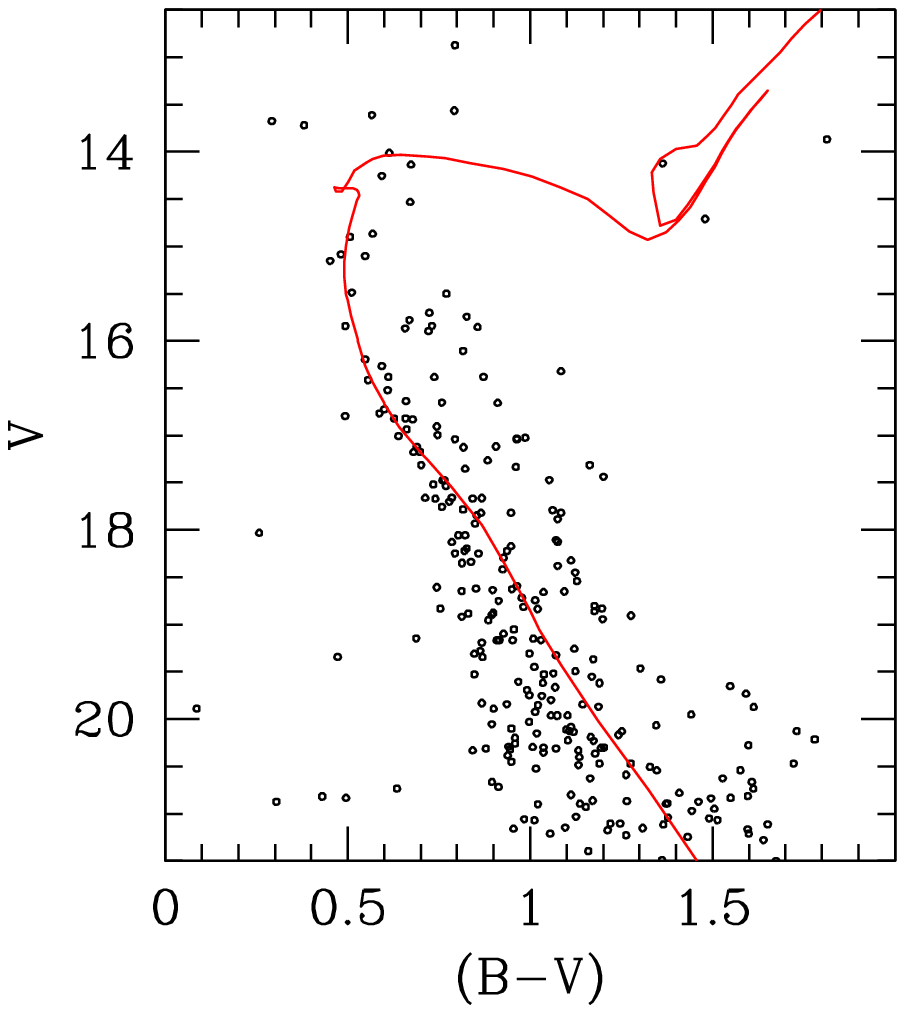}
   \caption{Isochrone solution for Haffner~4 in the V/B-V CMD. Fitting parameters are summarized in Table~4}
    \end{figure}

  \begin{figure}
   \centering
   \includegraphics[width=\columnwidth]{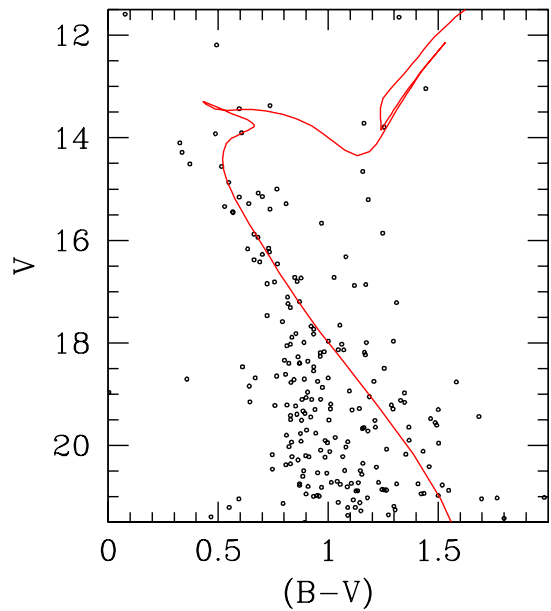}
   \caption{Isochrone solution for Ruprecht 10 in the V/B-V CMD. Fitting parameters are summarized in Table~4}
    \end{figure}

  \begin{figure}
   \centering
   \includegraphics[width=\columnwidth]{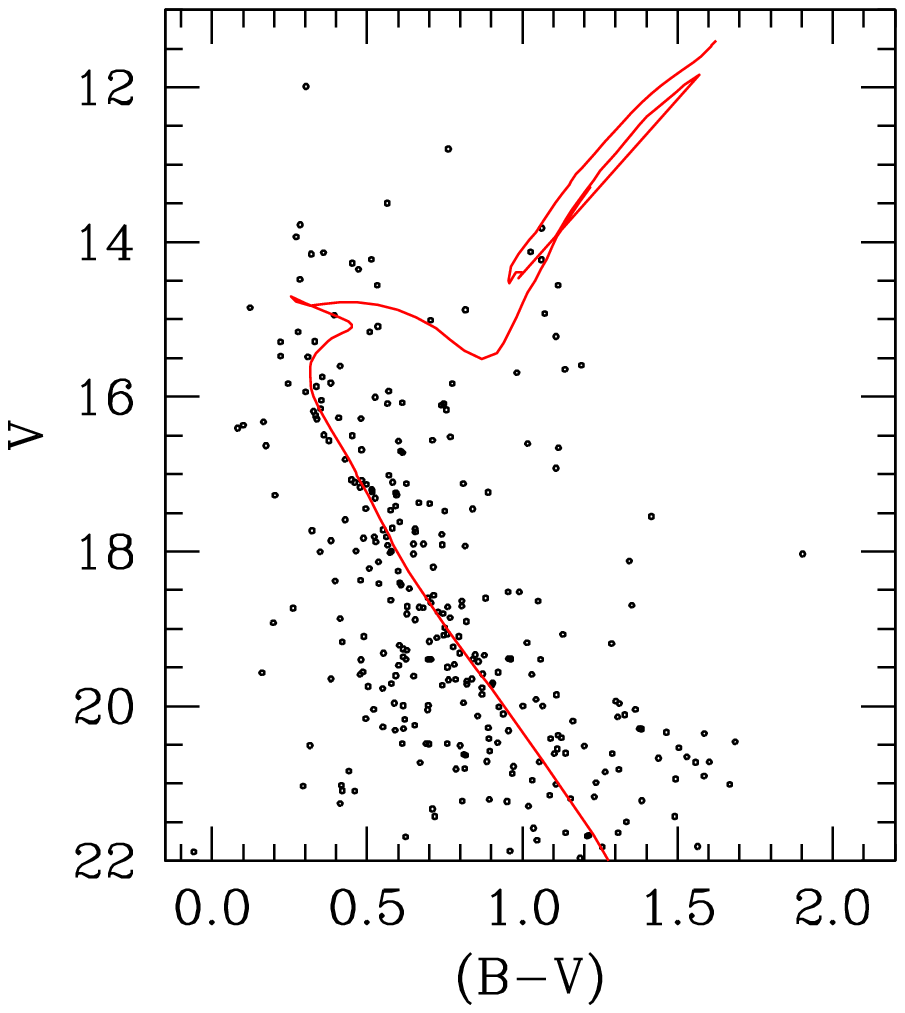}
   \caption{Isochrone solution for Haffner~7 in the V/B-V CMD. Fitting parameters are summarized in Table~4}
    \end{figure}

  \begin{figure}
   \centering
   \includegraphics[width=\columnwidth]{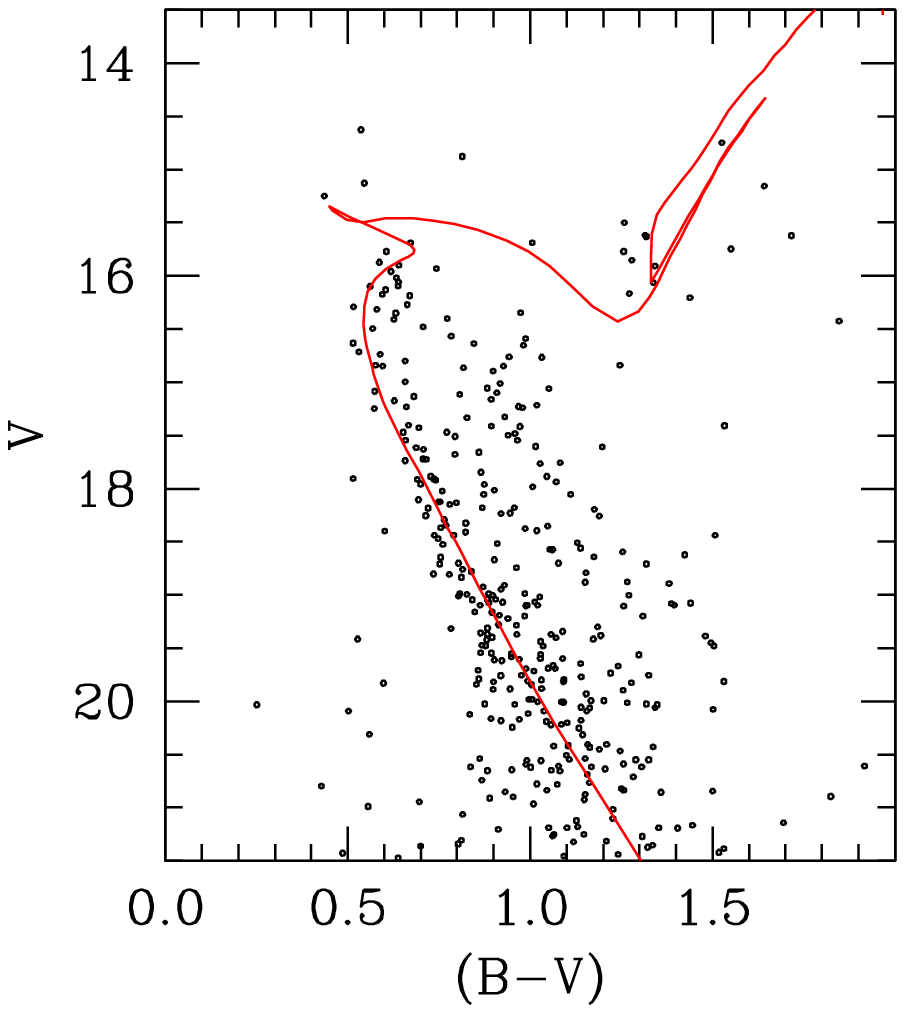}
   \caption{Isochrone solution for Haffner 11 in the V/B-V CMD. Fitting parameters are summarized in Table~4}
    \end{figure}

\noindent
{\bf Haffner~11: see Fig~17.}\\
This cluster results to be a nice intermediate-age open cluster, with a well defined
clump and not much contamination, when  stars inside the cluster radius are isolated,
and in agreement with all previous studies. It was quite an easy tasks
to super-impose on the cluster sequence a solar metallity isochrone for an age of 800 million years,
which yields a good fit in both the CMDs.
From this fit we derive a reddening E(B-V)=0.35$\pm$0.05, and an apparent distance modulus 
(m-M)$_V$ = 15.0$\pm$0.10. We therefore place the cluster at 6.0 kpc from the Sun, and 12.5
kpc from the Galactic center.\\

\noindent
Our results are therefore closer to Bica \& Bonatto (2005) than to Piatti et al. (2009).\\

\noindent
{\bf Haffner~15: see Figs~18 and 19.}\\
The preliminary estimate of the age of Haffner~15 in V\'azquez et al. (2008) was 600 million years,
significantly larger than the 15 Myr reported by Paunzen et al (2006). To solve this discrepancy
we make use of a different set of diagrams. In detail, we make use first of the U-B/B-V color-color
diagram, which is shown in Fig.~18. Here the solid line is an empirical Zero Age MS (ZAMS)
from Schmidt-Kaler (1982). 
The same ZAMS (dashed line) is shifted by E(B-V) = 1.05 along the reddening vector (indicated
by the arrow in the plot) to fit
the bulk of Haffner~15 stars. As justified above, we adopted a normal reddening law, which implies $R_V$ = 3.1
and E(U-B)/E(B-V) = 0.72.
To guide the eye, a few relevant spectral types are indicated, together with their displacement
along the reddening path.\\

\noindent
As a result,  the cluster reddening is E(B-V)=1.05$\pm$0.25, and the large
dispersion indicates the cluster suffers from  significant variable reddening.\\

\noindent
We now make use of the Q method (Strayzis 1991, Carraro et al. 2010) to derive individual stars reddening, and from them
compute their absolute magnitude and colors. 
These values are then used to build up the reddening-corrected CMDs in Fig.19, in the (B-V)$_o$ vs V$_o$ plane (left panel),
and in the (U-B)$_o$ vs V$_o$ plane (right panel). We super-pose to the star distribution the same empirical
ZAMS used before. Here the only free parameter is the distance modulus, which turns out to be $V_o-M_V$ = 13.00 $\pm$0.10,
in both diagrams. One can better appreciate the fit in the (U-B)$_o$ vs V$_o$ CMD, since the main sequence is more
tilted in this diagram. The TO point is located at V$_o$ $\sim$ 11.50 (V $\sim$ 14.5 ), which translates into M$_V$ = -1.50. This means
that the stars leaving the MS are of approximate spectral type B2, as confirmed also by inspecting the two color
diagram in Fig.~11. As a consequence, one can estimate the cluster age to be around 10-20 million years
(Marigo et al. 2008).\\

\noindent
The apparent distance modulus (m-M)$_V$ is then 16.0$\pm$0.10.
Overall, we therefore confirm Paunzen et al. (2006) results that Haffner~15 is indeed a young,
extremely reddened open cluster. It is located 3.5 kpc from the Sun, and 10.3 kpc from the Galactic center.
Such distance is compatible with Haffner~15 membership to the Perseus arm, or to the local (Orion) arm
extension in the third Galactic quadrant (Moitinho et al. 2006, Levine et al. 2006, V\'azquez et al. 2008).\\

  \begin{figure}
   \centering
   \includegraphics[width=\columnwidth]{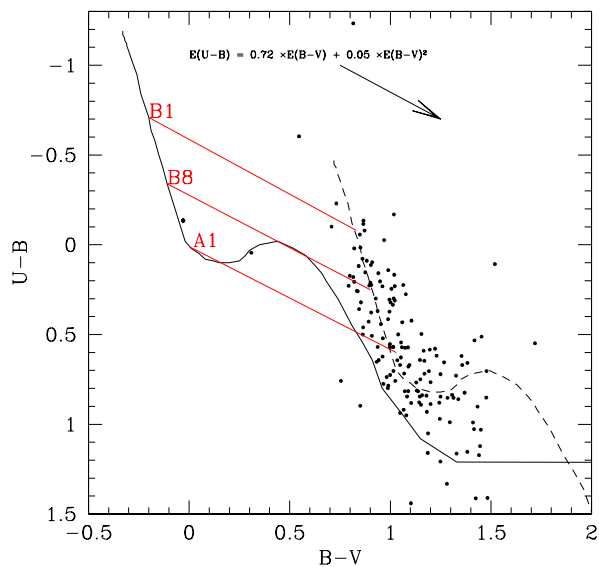}
   \caption{Two color diagrams for Haffner~15. Only stars falling inside the cluster radius
           are plotted. The solid line is an empirical unreddened ZAMS. 
           The same ZAMS, shifted by 
           E(B-V)=1.05 along the reddening path, is shown with a dashed line. The reddening 
           vector for a normal reddening law is indicated with an arrow. To guide the eye,
           a few relevant spectral types are also indicated along the zero reddening ZAMS,
           together with their displacement along the reddening vector direction.}
    \end{figure}

  \begin{figure}
   \centering
   \includegraphics[width=\columnwidth]{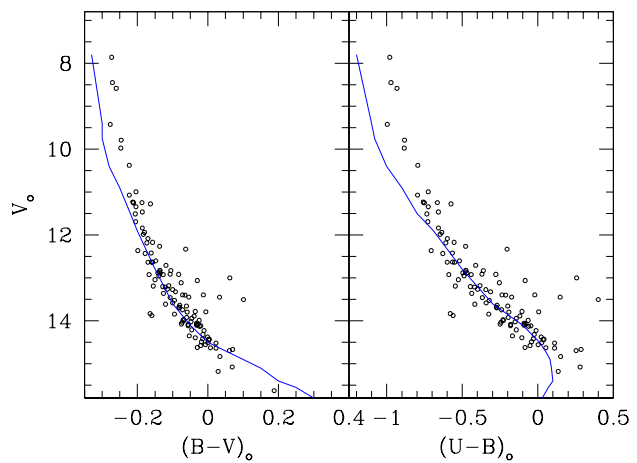}
   \caption{Reddening free CMDs in the (B-V) vs V plane (left panel), and (U-B) vs
   V plane (right panel). The solid line is a reddening free, empirical ZAMS shifted
   by $(V_O-M_V)$ = 13.00.}
    \end{figure}

\renewcommand{\thetable}{4}
\begin{table*}
\caption{Fundamental parameters estimated for the clusters.  $d_{\odot}$ indicates the distance
of a cluster from the Sun, Z the height below the Galactic plane, and $R_{GC}$ its distance to the Galactic center.} 
\begin{tabular}{rrrrrrrr}
\hline

Name &  Radius & E(B$-$V) & $(m-M)_V$ & $d_{\odot}$ & Age & Z & $R_{GC}$\\
 & arcmin & (mag) & (mag) & (kpc) & (Gyr) & (pc) & kpc \\
\hline
Berkeley~76   & 2.5$\pm$0.5 & 0.55$\pm$0.10 & 17.20$\pm$0.15 & 12.6 & $\sim$ 1.5  & $\sim$ -0.45 & $\sim$ 17.4\\
Haffner~4     & 2.5$\pm$0.5 & 0.50$\pm$0.10 & 14.80$\pm$0.15 &  4.5 & $\sim$ 0.5  & $\sim$ -0.28 & $\sim$ 11.9\\
Ruprecht~10   & 2.0$\pm$0.5 & 0.38$\pm$0.10 & 13.40$\pm$0.15 &  2.9 & $\sim$ 1.1  & $\sim$ -0.30 & $\sim$ 10.5\\
Haffner~7     & 1.5$\pm$0.5 & 0.13$\pm$0.10 & 13.65$\pm$0.15 &  4.5 & $\sim$ 1.5  & $\sim$ -0.50 & $\sim$ 11.3\\
Haffner~11    & 2.0$\pm$0.5 & 0.32$\pm$0.05 & 14.90$\pm$0.10 &  6.0 & $\sim$ 0.8  & $\sim$ -0.40 & $\sim$ 12.5\\
Haffner~15    & 2.0$\pm$0.5 & 1.05$\pm$0.25 & 16.00$\pm$0.10 &  3.5 & $\sim$ 0.02 & $\sim$ -0.25 & $\sim$ 10.3\\
\hline
\end{tabular}
\end{table*}

\section{Discussion and Conclusions}
We have presented and discussed UBVI CCD photometry of six Galactic star clusters projected
against the Canis Major overdensity in the third quadrant of the Milky Way:
Berkeley~76, Haffner~4, Ruprecht~10, Haffner~7, Haffner~11, and
Haffner~15. The fields
they are immersed in have already been analysed in V\'azquez et al (2008), where we searched for
young diffuse stellar populations to be used as tracers of the spiral structures of the outer disk.
Here we concentrated on the clusters themselves, with the goal of deriving 
estimates of their age and distance.\\

\noindent
We can summarize our results as follows (see also Table~4):

\begin{description}

\item $\bullet$ according to star counts, all the objects appear as significant overdensities with respect to the field;

\item $\bullet$ the cluster Haffner~15 is the only young object of the sample. For its age and distance it
is a probable member of the Perseus arm or of the extension of the local arm into the third Galactic
quadrant. This cluster is heavily and differentialy reddened; 

\item $\bullet$ we found one old star cluster at the extreme periphery of the disk, at a mean distance of more
than 17 kpc from the Galactic center: Berkeley~76; only Saurer~1 and Berkeley~29 are located further
than them (Carraro \& Baume 2003);

\item $\bullet$ Ruprecht~10 and Haffner~7,  two clusters that were never studied before, 
 are  found to have ages larger than 1 Gyr;

\item $\bullet$ the spatial shape of these clusters, as highlighted with star counts, 
may witness their dynamical status. Haffner~15, the youngest, has an almost 2-dimensional
circular shape, while all the others -older than the Hyades- have complicated structures, very far from spherical. 
We suggest that they are  close to dissolution and merging with the general Galactic field, although their
actual dynamical status can only be confirmed with invidual stars kinematics.

\end{description}

\noindent
We believe this sample of clusters is quite promising, and deserves a much closer look in the future,
which  unfortunately GAIA will not be able to provide at those distances.
The oldest, most distant clusters, in particular,  are excellent targets to further probe the slope of the abundance gradient
in the outer disk, and its evolution with time (Twarog et al. 1997, Carraro et al. 2007, Magrini et al. 2009).

\section{Acknowledgment}
G. Carraro would like to mention and acknowledge the great
support from Cerro Tololo Observatory staff, in particular from Edgardo Cosgrove.
We thank warmly Sandy Strunk for
reading the manuscript and helping us to improve the language.
This study made use of the SIMBAD and WEBDA databases.

\end{document}